\documentstyle[12pt,a41,epsfig,cite,amssymb]{article}

\newcommand{\bea}{\begin{eqnarray}}
\newcommand{\bq}{\begin{equation}}
\newcommand{\eea}{\end{eqnarray}}
\newcommand{\eq}{\end{equation}}

\newcommand{\lsim}{\raisebox{-0.07cm   }
{$\, \stackrel{<}{{\scriptstyle\sim}}\, $}}

\newcommand\GeV{\,\mbox{GeV}}
\newcommand\TeV{\,\mbox{TeV}}

\newcommand\non{\nonumber}
\newcommand\del{\delta}
\newcommand\be{\begin{eqnarray}}
\newcommand\ee{\end{eqnarray}}
\newcommand\Sa{{\rm S}}

\newcommand\Li{{\rm Li}}
\newcommand{\Sf}{{\rm S}_{1,2}}
\newcommand\SS{{\rm S}}
\newcommand\PV{\, \mbox{\boldmath $P$}}
\newcommand\DV{\, \mbox{\boldmath $D$}}
\newcommand\EV{\, \mbox{\boldmath $E$}}

\newcommand\RV{\, \mbox{\boldmath $R$}}
\newcommand\M{\, \mbox{\boldmath $M$}}
\begin{document}
\noindent
\sloppy
\thispagestyle{empty}
\begin{flushleft}
DESY 04-064 \hfill
{\tt hep-ph/0701019}\\
SFB/CPP--06--63\\
December 2006
\end{flushleft}
%
\vspace*{\fill}
\begin{center}
{\LARGE\bf Universal Higher Order Singlet QED}

\vspace{2mm}
{\LARGE\bf Corrections to Unpolarized Lepton Scattering}

\vspace{2cm}
\large
Johannes Bl\"umlein~$^a$  and
Hiroyuki Kawamura~$^{a,b}$
\\
\vspace{2em}
\normalsize
{\it $^a$~Deutsches Elektronen--Synchrotron, DESY,\\
Platanenallee 6, D--15738 Zeuthen, Germany}
\\
{\it $^b$~Radiation Laboratory, RIKEN, Wako 351-0198,  Japan}
\\
\vspace{2em}
\end{center}
\vspace*{\fill}
%
\begin{abstract}
\noindent
We calculate the universal flavor-singlet radiative QED corrections to 
unpolarized lepton scattering applicable to general differential scattering 
cross sections, involving charged fermions or photons in initial or final 
states. The radiators are derived  to $O((\alpha \ln(Q^2/m_f^2))^5)$ in analytic 
form. Numerical illustrations are given.
\end{abstract}
\vspace*{\fill}
\newpage
\section{Introduction}
\label{sec-1}

\vspace{1mm}
\noindent
QED corrections to integral and differential cross sections of light 
charged lepton--anti-lepton scattering or deeply inelastic lepton--nucleon
scattering turn out to be quite large in some kinematic
regions~\cite{EXAMP1,Beenakker:1989km,Blumlein:1989gk,Bardin:1996ch}. 
This applies in particular to 
the Bremsstrahlung contributions due to significant shifts in the 
kinematics of the underlying differential scattering cross sections.
The universal corrections  can be grouped into flavor non--singlet and 
flavor singlet contributions. In the orders $O((e_f^2 \alpha L)^k)$, with 
$\alpha$ the fine structure constant, $L = \ln(Q^2/m_f^2)$,  $Q^2$ 
the typical virtuality of the process and $m_f, e_f$ the fermion 
mass and charge, respectively, the non--singlet contributions stem from 
the leading order anomalous dimension in QED, 
$P_{ff}(x,Q^2)$. The diagonal elements of the singlet anomalous dimension matrix 
contain a $\delta(1-x)$-distribution 
and are distribution-valued due to  $(...)_+$-distributions. Therefore the  
numerical Mellin-inversion does badly converge in the region of 
$x \lessapprox 1$.
Analytic representations are required to high 
enough order $k$ in the fine structure constant $\alpha$ to cover all universal 
effects for the energy ranges probed at present day colliders and those
to be built in the foreseeable future. 
This applies to high luminosity experiments at future linear colliders \cite{ILC} and as well to 
the search for rare reactions at LHC.
The second order universal corrections for 
various processes are known for a long time 
\cite{Berends:1987ab,Altarelli:1986kq,Blumlein:1994ii,Arbuzov:1995id}. 
The 3rd order corrections were given in 
\cite{Jezabek:1992bx}~\footnote{For an application to the $Z$--peak see \cite{Montagna:1996jv}.} 
for the flavor non-singlet and in 
\cite{Skrzypek:1992vk} also for the singlet case. Later   
the 5th order non-singlet corrections were given in 
\cite{Przybycien:1993qe} and recalculated in 
\cite{Arbuzov:1999cq}
and \cite{BLKA,Blumlein:2004bs}~\footnote{The results of \cite{Przybycien:1993qe} and  \cite{Blumlein:2004bs} 
agree but partly disagree in the 5th order with  \cite{Arbuzov:1999cq}.}. 
In Ref.~\cite{Blumlein:2004bs} a very 
compact form was given for the non-singlet contributions, which are the 
same for polarized and unpolarized scattering. There also the polarized 
singlet contributions were calculated to $O((\alpha L)^5)$.
In the present paper the unpolarized singlet evolution kernels are calculated 
to $O((\alpha L)^5)$ which supplements earlier investigations for the
flavor non--singlet kernel \cite{Blumlein:2004bs}. A second class of universal
QED corrections was treated previously in the non-singlet \cite{Blumlein:1998yz} 
and polarized singlet case \cite{Blumlein:2004bs}.
It concerns the leading order small $x$ resummations of $O((\alpha \ln^2(x))^k)$. 
These resummations are based on 
corresponding resummations in QCD~\cite{SX1}. In the unpolarized (pure) singlet case 
the leading order small-$x$ QCD resummations \cite{Fadin:1975cb} result 
from the non-abelian gluon coupling, which is absent in  QED. 
Therefore a transformation of the respective kernels is not possible in this case.

The paper is organized as follows. The general framework to derive the radiators
in $O((\alpha L)^k)$ is outlined in section~2. In section~3 the leading order 
singlet radiators are calculated to $O((\alpha L)^5)$. In section~4 numerical
illustrations are given and section~5 contains the conclusions. An appendix lists
useful convolution relations which were needed in the calculation and are of use
in other QED and QCD calculations.

\section{The Solution of Singlet Evolution Equations}

\vspace{1mm}
\noindent
The universal QED corrections $O((\alpha L)^k)$ for general values of the collinear 
radiation momentum fraction $x$ can be expressed solving the singlet evolution 
equations starting at a low scale $Q_0^2$. This scale may be identified with a typical 
charged lepton mass $m_f$ squared. The running 
coupling constant $a(Q^2) = \alpha(Q^2)/(4\pi)$  obeys the evolution equation
\begin{eqnarray}
\frac{d a(Q^2)}{d\ln(Q^2)} = - \sum_{k=0}^{\infty} \beta_k a^{k+2}(Q^2)~,
\end{eqnarray}
where $\beta_k$ denote the expansion coefficients of the QED $\beta$-function in the
${\overline{\rm MS}}$-scheme, $\beta_0 = -4/3 N_f, \beta_1 = -4 N_f, \beta_2 = 2 N_f 
+(44/9) N_f^2$ etc.~~\cite{TAR} for $N_f$ active charged lepton species. 
At leading order the solution
\begin{eqnarray}
a(Q^2) = \frac{a(m_f^2)}{1- \frac{4}{3} N_f a(m_f^2) L}~,
\end{eqnarray}
with $L = \ln(Q^2/m_f^2)$ is obtained.

The universal radiators are found as solutions of the leading order QED renormalization group
equations associated to the collinear singularities. The following QED--distributions are
of relevance~: 
\begin{eqnarray}
D_{\rm NS}^f(a,x) &=& D^f(a,x) - D^{\overline{f}}(a,x)\\
D_{\rm \Sigma}^f(a,x) &=& D^f(a,x) + D^{\overline{f}}(a,x)\\
D_{\gamma \gamma}(a,x) &=& D_{22}(a,x)\\
D_{\gamma f}(a,x) &=& D_{21}(a,x)\\
D_{f \gamma}(a,x) &=& D_{12}(a,x)
\end{eqnarray}
The non-singlet distribution $D_{\rm NS}^f(a,x)$ was dealt with in \cite{Przybycien:1993qe,
Arbuzov:1999cq,Blumlein:2004bs} before. Here the  $D_{ij}(a,x)$ denote the respective
matrix element of the singlet radiator $\DV_S(a,x)$ given below.

The singlet radiator functions at the scale $Q_0^2$ are 
\begin{eqnarray}
\DV(Q_0^2)(x) \equiv \DV(a_0) = {\bf 1}~~\delta(1-x)
\end{eqnarray}
since both the charged leptons and the photon are considered to be asymptotically stable
particles. The evolution equations read
\begin{eqnarray}
\label{eqEVO1}
\frac{\partial \DV_S(a,x)}{\partial a}
&=&
- \frac{1}{a}
\frac{\sum_{k=0}^{\infty} a^k \PV_k(x)}
{\sum_{k=0}^{\infty} a^k \beta_k} \otimes \DV_S(a,x) 
\nonumber\\
&=& - \frac{1}{\beta_0 a} \Biggl[\PV_0(x) + a\left(\PV_1(x)
- \frac{\beta_1}{\beta_0} \PV_0(x)\right) + O(a^2)\Biggr]
\otimes \DV_S(a,x)~. 
\end{eqnarray}
The Mellin convolution is defined by
\begin{eqnarray}
\label{melcon}
A(x) \otimes B(x) = \int_0^1 d x_1 \int_0^1 d x_2 \delta(x-x_1 x_2) A(x_1) B(x_2)~.
\end{eqnarray}
The Mellin transform
\begin{eqnarray}
\M[f(x)](N) = \int_0^1 dx~x^{N-1} f(x)
\end{eqnarray}
diagonalizes the convolution (\ref{melcon}) to
\begin{eqnarray}
\M[A(x) \otimes B(x)](N) = \M[A(x)] (N) \cdot \M[B(x)](N)~.
\end{eqnarray}
In some of the radiators $(...)_+$ distributions emerge, which are defined relative to 
the set of smooth test functions $\phi(x)$ with compact support by
\begin{eqnarray}
\int_0^1 dx \left[F(x)\right]_+ \phi(x) = \int_0^1 dx F(x) \left[\phi(x)-\phi(1)\right]~.
\end{eqnarray}

Eq.~(\ref{eqEVO1}) may be solved easiest in Mellin space as a matrix-valued ordinary 
differential
equation to all orders, see e.g. \cite{Furmanski:1981cw,Blumlein:1997em}. 
The leading order solution reads
\begin{eqnarray}
\label{eqSLO}
\DV_{S,0}(a,x) &=& \left[\exp(-\RV_0(x) \ln(a/a_0)
)_\otimes\right] \otimes \DV_S(a_0,x) \equiv \EV_0(a,a_0,x) \otimes
\DV_S(a_0,x)~,
\end{eqnarray}
where $\RV_0= \PV_0/\beta_0$ and $\EV_0$ denotes the leading order singlet evolution 
operator. We use the short--hand notation
\begin{eqnarray}
\left[f(g(x))_\otimes\right] = \sum_{k=0}^{\infty} \frac{f^{(k)}(0)}{k!}
\otimes_{l=1}^k g(x)~,
\end{eqnarray}
with $\otimes_{l=1}^k$ the $k$-fold convolution. 
The singlet solution (\ref{eqSLO}) to $k$th order in $\alpha(Q^2) L$ therefore requires
to calculate $k$--fold convolutions of the leading order matrix of splitting functions.

The method described above can be extended to sub-leading logarithmic contributions,
i.e. terms of $O(\alpha^2 L)$ etc. These contributions contain process-dependent parts
being described by Wilson coefficients in inclusive situations or the corresponding
semi-inclusive quantities. These corrections are neither universal nor independent of the
measurement chosen for the kinematic variables. Examples are the $O(\alpha^2 L)$  corrections
for the initial state radiation in $e^+e^-$ annihilation \cite{Berends:1987ab},
the $O(\alpha^2 L)$ initial and final state radiation corrections to deeply inelastic 
scattering 
\cite{Blumlein:2002fy} and the the $O(\alpha^2 \ln(m_{\mu}/m_e))$ corrections to the
electron spectrum in muon decay \cite{Arbuzov:2002cn}.
 
\section{The Leading Order Solution to \mbox{\boldmath $O[(\alpha L)^5]$}}

\vspace{1mm}
\noindent
In the following we derive the solution for the unpolarized singlet QED 
evolution kernels up to $O((\alpha L)^5)$. The singlet evolution equation 
is solved in the running coupling $a(Q^2)$. However, one may re-parameterize
the representation and express the evolution kernel directly in terms of
$a_0 = a(m_f^2)$ by
\begin{eqnarray}
\label{EV0}
\EV_0(a,x) &=& {\bf 1}~
\delta(1-x) + \PV_0(x) a_0 L + \left( \frac{1}{2}
\PV_0^{(1)}(x) + \frac{2}{3} \PV_0(x)\right) (a_0 L)^2          \\ & &
+\left(\frac{1}{6} \PV_0^{(2)}(x)
      +\frac{2}{3} \PV_0^{(1)}(x)
      +\frac{16}{27} \PV_0(x)\right) (a_0 L)^3 \nonumber\\ & &
+\left(\frac{1}{24} \PV_0^{(3)}(x)
      +\frac{1}{3} \PV_0^{(2)}(x)
      +\frac{22}{27} \PV_0^{(1)}(x)
      +\frac{16}{27} \PV_0(x)\right) (a_0 L)^4 \nonumber\\ & &
+\left(\frac{1}{120} \PV_0^{(4)}(x)
      +\frac{1}{9} \PV_0^{(3)}(x)
      +\frac{14}{27} \PV_0^{(2)}(x)
      +\frac{80}{81} \PV_0^{(1)}(x)
      +\frac{256}{405} \PV_0(x)\right) (a_0 L)^5~. \nonumber
\end{eqnarray}
Here the matrices $\PV_0^{(k)}$ are
\begin{eqnarray}
\PV_0^{(k)}(x) = \otimes^{(k)} \PV_0(x)~,
\end{eqnarray}
i.e. $ \otimes^{(1)} \PV_0(x) =\left[\PV_0 \otimes \PV_0\right](x)$ etc.
The corresponding expressions for the $(1,1)$--components of
$\PV_0^{(k)}(x)$ are given relative to the non--singlet components
$\PV_{\rm NS}^{(k)}(x)$, see \cite{Blumlein:2004bs}. 
The singlet--matrices are 
\begin{eqnarray}
\PV_0^{(k)}(x) = \left(\begin{array}{cc} 
                                     P_{11}^{(k+1)}(x) & P_{12}^{(k+1)}(x) 
\\
                                     P_{21}^{(k+1)}(x) & P_{22}^{(k+1)}(x) 
\end{array} \right)~,
\end{eqnarray}
with  components $P_{ij}^{(k)}(x)$ given below in Eqs.~(\ref{eqSP0}--\ref{eqSP2}). 
They were calculated using the convolution formulae of 
Appendix~A and relations given in \cite{Blumlein:2004bs,Blumlein:2000wh} before. 
The projections $P_{ij}^{(k)}$ describe the splitting of a fermion into 
a fermion (1,1), of a photon into a fermion, positron, respectively
(1,2), a fermion into
a photon (2,1), and a photon into a photon (2,2) in $k$th order in the 
renormalized coupling constant. 

The leading order QED splitting functions can be obtained identifying $T_R 
= C_F = 1$
and $C_A = 0$ in the QCD splitting functions \cite{Gross:1974cs,Altarelli:1977zs} 
in accordance with 
the gauge group $U(1)$. 

\vspace{3mm}
{\sf 1st order terms:}

\vspace{1mm}
\begin{eqnarray}
\label{eqSP0}
P_{11}^{(1)}(x)&=& P_{11 \rm NS}^{(1)}(x) + P_{11 \rm PS}^{(1)}(x) 
= \frac{2}{(1-x)_+}-1-x+\frac{3}{2}\delta(1-x)\\
P_{11 \rm PS}^{(1)}(x) &=& 0 \\
P_{12}^{(1)}(x)&=&2[x^2+(1-x)^2]=4x^2-4x+2\\
P_{21}^{(1)}(x)&=&\frac{1+(1-x)^2}{x}=x-2+\frac{2}{x}\\
\label{e22-1}
P_{22}^{(2)}(x)&=&-\frac{2}{3}\delta(1-x)
\end{eqnarray}
The $\delta(1-x)$ distribution in (\ref{e22-1}) emerges due to momentum
conservation for the photon momentum 
$\int_0^1 dx x\left[P_{12}(x) + P_{22}(x)\right] = 1$.

\vspace{3mm}
The {\sf 2nd order terms} are :

\vspace{1mm}
\begin{eqnarray}
P_{11}^{(2)}(x) &=& P_{11 \rm NS}^{(2)}(x) + P_{11 \rm PS}^{(2)}(x) 
\non\\
&=&8\left(\frac{\ln(1-x)}{1-x}\right)_+ 
-4(1+x)\ln(1-x) +\ln (x)\left[7+7x-\frac{4}{1-x}\right] \non\\ 
&& 
+\frac{6}{(1-x)_+}-3-3x-\frac{8}{3}x^2+\frac{8}{3x} 
+\left[\frac{9}{4}-4\zeta(2)\right]\del(1-x) \\ 
P_{12}^{(2)}(x) &=& 4(1-2x+2x^2)\ln(1-x)-2(1-2x+4x^2)\ln(x) 
-\frac{7}{3}+\frac{20}{3}x-\frac{8}{3}x^2 \\ 
P_{21}^{(2)}(x)&=& 2\ln(1-x)\left[x-2+\frac{2}{x}\right]
+(2-x)\ln(x) +\frac{10}{3}-\frac{7}{6}x-\frac{4}{3x} \\ 
P_{22}^{(2)}(x)&=&4(1+x)\ln(x) +2-2x-\frac{8}{3}x^2+\frac{8}{3x} 
+\frac{4}{9}\del(1-x) 
\end{eqnarray} 
They were derived in \cite{Altarelli:1986kq} and various other places before.

\vspace{3mm}
{\sf 3rd order terms:}

\vspace{1mm}
\begin{eqnarray}
P_{11}^{(3)}(x) &=& P_{11 \rm NS}^{(3)}(x) + P_{11 \rm PS}^{(3)}(x) 
\non\\ &=&
24\left(\frac{\ln^2(1-x)}{1 -x}\right)_{+}
-12(1+x)\ln^2(1-x) +36\left(\frac{\ln(1-x)}{1-x}\right)_{+} \non\\
&& 
-\ln(1-x)\left[22+14x+\frac{32}{3}x^2-\frac{32}{3x}\right] -24\ln(x)\frac{\ln(1-x)}{1-x} 
\non\\
&& 
+34(1+x)\ln(x) \ln(1-x) +\ln^2(x) \left[\frac{4}{1-x}-\frac{15}{2}-\frac{15}{2}x\right] 
\non\\
&& 
+\ln(x)\left[\frac{83}{6}+\frac{59}{6}x+\frac{32}{3}x^2 -\frac{18}{1-x}\right]+22(1+x)\Li_2(1-x) 
\non\\&& 
+\left[\frac{27}{2}-24\zeta(2)\right]\frac{1}{(1-x)_+} 
-\frac{317}{12}+\frac{155}{12}x+\frac{16}{9}x^2 
-\frac{16}{9x}+12(1+x)\zeta(2) \non\\&& 
+\left[\frac{27}{8}-18\zeta(2)+16\zeta(3)\right]\del(1-x)\\ 
P_{12}^{(3)}(x) &=& 
8(1-2x+2x^2)\ln^2(1-x)-\frac{4}{3}(5-16x+4x^2)\ln(1-x) 
\non\\&& 
-8(1-2x+4x^2)\ln(x)\ln(1-x)-(3-6x-8x^2)\ln^2(x) 
\non\\&& -\ln(x)\left[\frac{32}{3}+\frac{26}{3}x-16x^2\right] 
-16x^2\Li_2(1-x) 
\non\\ &&
-\frac{521}{18}+\frac{491}{9}x-\frac{232}{9}x^2+\frac{32}{9x}
-8(1-2x+2x^2)\zeta(2) \\ 
P_{21}^{(3)}(x) &=& 
4\left[x-2+\frac{2}{x}\right]\ln^2(1-x) -\frac{2}{3}\left[5x-16+\frac{4}{x}\right]\ln(1-x) \non\\&& 
+4(2-x)\ln(x)\ln(1-x)+\frac{3}{2}(2-x)\ln^2(x) +\frac{1}{3}(26x-19-\frac{16}{x})\ln(x) \non\\&& 
+\frac{8}{x}\Li_2(1-x)-4\left[x-2+\frac{2}{x}\right]\zeta(2) 
+\frac{491}{18}-\frac{521}{36}x+\frac{16}{9}x^2-\frac{116}{9x} \\ 
P_{22}^{(3)}(x)&=& 
4\ln(1-x)\left[1-x-\frac{4}{3}x^2+\frac{4}{3}\frac{1}{x}\right] +8(1+x)\ln(x)\ln(1-x)-2(1+x)\ln^2(x) 
\non\\&& 
-\frac{4}{3}(4+x-4x^2)\ln(x) +8(1+x)\Li_2(1-x) -\frac{31}{3}(1-x)+\frac{32}{9}\left[x^2-\frac{1}{x}\right] 
\non\\&& 
-\frac{8}{27}\del(1-x) 
\end{eqnarray} 

\vspace{3mm}
{\sf 4th order terms:}

\vspace{1mm}
\noindent
Here and for the 5th order terms we refer to the expressions $P_{\rm NS}^{(k)}(x)$
given in Ref.~\cite{Blumlein:2004bs} for brevity.
\begin{eqnarray} P_{11}^{(4)}(x) &=&  
P_{\rm NS}^{(4)}(x) +
48(1+x)\ln(x)\ln^2(1-x)-24(1+x)\ln^2(x)\ln(1-x) \non\\&& 
-\frac{2}{3}(1+x)\ln^3(x) 
+8\left[3-3x-4x^2+\frac{4}{x}\right]\ln^2(1-x) \non\\&& - 
8\ln(x)\ln(1-x)\left[\frac{4}{3}-\frac{14}{3}x-8x^2\right] + 
\ln^2(x)\left[-\frac{25}{3}+\frac{5}{3}x-16x^2\right] \non\\&& - 
4\left[\frac{73}{3}-\frac{73}{3}x-\frac{16}{9}x^2
                        +\frac{16}{9x}\right]\ln(1-x)
\non\\&&
-\left[\frac{569}{9}+\frac{1445}{9}x+\frac{128}{9}x^2
+\frac{64}{9x}+48\zeta(2)+48\zeta(2)x\right]\ln(x)
\non\\&&
+ 96(1+x)\ln(1-x)\Li_2(1-x)
+8\left[\frac{5}{3}+\frac{5}{3}x+4x^2+\frac{4}{x}\right]\Li_2(1-x)
\non\\&&       
- 96(1+x)\Li_3(1-x) + 48(1+x)\Sf(1-x)
-\frac{745}{18}+\frac{745}{18}x+\frac{224}{9}x^2
-\frac{224}{9x}
\non\\&&
- 8\left[3-3x-4x^2+\frac{4}{x}\right]\zeta(2)
\\
P_{12}^{(4)}&=&
16(1-2x+2x^2)\ln^3(1-x)-24(1-2x+4x^2)\ln(x)\ln^2(1-x)
\non\\&&
-2\ln^2(x)\ln(1-x)(5-10x-24x^2)
+ \ln^3(x)\left[\frac{7}{3}-\frac{14}{3}x-\frac{16}{3}x^2\right]
\non\\&&
-4\ln^2(1-x)\left[\frac{13}{3}-\frac{44}{3}x+\frac{8}{3}x^2\right]
- 4\ln(x)\ln(1-x)\left[\frac{32}{3}+\frac{35}{3}x-16x^2\right]
\non\\&&       
+ \ln^2(x)\left[\frac{79}{6}+\frac{14}{3}x-\frac{80}{3}x^2\right]
\non\\&&
- \ln(1-x)\left[\frac{959}{9}-\frac{1978}{9}x+\frac{320}{3}x^2
-\frac{128}{9x} + 48\zeta(2)- 96\zeta(2)x +96\zeta(2)x^2\right]
\non\\&&
+\ln(x)\left[\frac{1505}{18}-\frac{614}{9}x+\frac{832}{9}x^2 
            +24\zeta(2)-48\zeta(2)x+96\zeta(2)x^2\right]
\non\\&&
-96x^2 \ln(1-x)\Li_2(1-x)-44(1-2x)\ln(x)\Li_2(1-x)
\non\\&&
-4\Li_2(1-x)\left[15-3x-\frac{40}{3}x^2\right]+ 96x^2\Li_3(1-x)
\non\\&&
-4(13-26x+16x^2)\Sf(1-x)
+\frac{18065}{108} -\frac{5947}{27}x 
+ \frac{560}{9}x^2 -\frac{128}{27x}
\non\\&&
+4\left[\frac{13}{3}-\frac{44}{3}+\frac{8}{3}x^2\right]\zeta(2) 
+ 32(1-2x+2x^2)\zeta(3)
\\
P_{21}^{(4)}&=&
8\left[x-2+\frac{2}{x}\right]\ln^3(1-x)
+12(2-x)\ln(x)\ln^2(1-x)
\non\\&&
+5(2-x)\ln^2(x)\ln(1-x)-\frac{7}{6}(2-x)\ln^3(x)
+\frac{2}{3}
\ln^2(1-x)\left[44-13x-\frac{8}{x}\right]
\non\\&&
-2\ln(x)\ln(1-x)
\left[\frac{53}{3}-\frac{58}{3}x+\frac{32}{3}\frac{1}{x}\right]
+\ln^2(x)\left[\frac{16}{3}-\frac{101}{12}x\right]
\non\\&&
+\ln(1-x)\left[\frac{989}{9}-\frac{959}{18}x+\frac{64}{9}x^2
-\frac{160}{3x}+48\zeta(2)-24\zeta(2)x
-\frac{48}{x}\zeta(2)\right]
\non\\&&      
+\ln(x)\left[-\frac{682}{9}+\frac{413}{36}x-\frac{64}{9}x^2
                 +\frac{64}{9x}-24\zeta(2)+12\zeta(2)x \right]
\non\\&&     
+\frac{48}{x}\ln(1-x)\Li_2(1-x)+22(2-x)\ln(x)\Li_2(1-x)
\non\\&&
-2\Li_2(1-x)\left[3-15x+\frac{40}{3x}\right]
-\frac{48}{x}\Li_3(1-x)+ 2\left[26-13x+\frac{8}{x}\right]\Sf(1-x)
\non\\&&
- \frac{5947}{54} + \frac{18065}{216}x 
- \frac{64}{27}x^2 + \frac{280}{9x}
- \frac{2}{3}\left[44-13x-\frac{8}{x}\right]\zeta(2)
\non\\&&
- 16\left[2-x-\frac{2}{x}\right]\zeta(3)
\\
P_{22}^{(4)}(x) &=&
16(1+x)\ln(x)\ln^2(1-x)-8(1+x)\ln^2(x)\ln(1-x)-2(1+x)\ln^3(x)
\nonumber\\&&
+8\left[1-x-\frac{4}{3}x^2+\frac{4}{3x}\right]\ln^2(1-x)
-\frac{16}{3}(2-x-4x^2)\ln(x)\ln(1-x)
\nonumber\\&&
-\frac{1}{3}(19-23x+16x^2)\ln^2(x)
- 4\ln(1-x)\left[9-9x-\frac{16}{9}x^2+\frac{16}{9x}\right]
\nonumber\\&&
-\ln(x)\left[\frac{217}{3}+\frac{325}{3}x+\frac{128}{9}x^2
+\frac{64}{9x}+16\zeta(2)+16\zeta(2)x\right]
\nonumber\\&&
+32(1+x)\ln(1-x)\Li_2(1-x)
-\frac{8}{3}\Li_2(1-x)\left[1+x-4x^2-\frac{4}{x}\right]
\nonumber\\&&
-32(1+x)\Li_3(1-x)+16(1+x)\Sf(1-x)
\nonumber\\&&
-\frac{1331}{18}+\frac{1331}{18}x+\frac{608}{27}x^2
-\frac{608}{27x}
-8\left[1-x-\frac{4}{3}x^2+\frac{4}{3x}\right]\zeta(2)
\nonumber\\&& +\frac{16}{81}\delta(1-x) 
\end{eqnarray}

\vspace{3mm}
{\sf 5th order terms:}

\vspace{1mm}
\begin{eqnarray}
P_{11}^{(5)}(x) & = &  
P_{\rm NS}^{(5)}(x) +
128(1+x)\ln(x)\ln^3(1-x)-96(1+x)\ln^2(x)\ln^2(1-x)
\non\\&&
+\frac{4}{3}(1+x)\ln^4(x)
+64\left[1-x-\frac{4}{3}x^2+\frac{4}{3x}\right]\ln^3(1-x)
\non\\&&
-32(1-5x-8x^2)\ln(x)\ln^2(1-x)-8(7+x+16x^2)\ln^2(x)\ln(1-x)
\non\\&&
+\frac{8}{9}\left[10-5x+16x^2\right]\ln^3(x)
-64\left[6-6x-\frac{x^2}{3}+\frac{1}{3x}\right]\ln^2(1-x)
\non\\&&
-32\ln(x)\ln(1-x)\left[\frac{97}{9}+\frac{313}{9}x
           +\frac{8}{3}x^2+\frac{4}{3x}
                                   +12\zeta(2)+12\zeta(2)x\right]
\non\\&&       
+16\ln^2(x)
\left[\frac{83}{9}+\frac{122}{9}x+2x^2+6\zeta(2)+6\zeta(2)x\right]
\non\\&&
-4\ln(1-x)\left[\frac{331}{9}-\frac{331}{9}x-\frac{1024}{27}x^2
      +\frac{1024}{27x}\right.
\non\\&&\left.\hspace{4cm}
+48\zeta(2)-48\zeta(2)x-64\zeta(2)x^2+\frac{64}{x}\zeta(2)\right]
\non\\&&       
+4\ln(x)\left[\frac{1990}{27}+\frac{997}{27}x-\frac{320}{9}x^2
+\frac{64}{27x}\right.
\non\\&&\left.\hspace{4cm}
+ 8\zeta(2)-40\zeta(2)x-64\zeta(2)x^2+64\zeta(3)+64\zeta(3)x\right]
\non\\&&       
+ 384(1+x)\ln^2(1-x)\Li_2(1-x)-96(1+x)\ln^2(x)\Li_2(1-x)
\non\\&&
+128\left[1+x+2x^2+\frac{2}{x}\right]\ln(1-x)\Li_2(1-x)
\non\\&&
-144(1-x)\ln(x)\Li_2(1-x)-768(1+x)\ln(1-x)\Li_3(1-x)
\non\\&&       
+ 384(1+x)\ln(1-x)\Sf(1-x)-256(1+x)\ln(x)\Sf(1-x)
\non\\&&
-32\left[\frac{205}{9}+\frac{205}{9}x+2x^2+\frac{2}{x}
                         +12\zeta(2)+12\zeta(2)x\right]\Li_2(1-x)
\non\\&&       
-128\left[1+x+2x^2+\frac{2}{x}\right]\Li_3(1-x)
-16\left[7-15x-\frac{32}{3}x^2-\frac{16}{3x}\right]\Sf(1-x)
\non\\&&       
+768(1+x)\Li_4(1-x)-192(1+x)\SS_{1,3}(1-x)-384(1+x)\SS_{2,2}(1-x)
\non\\&&
+\frac{2}{27}\left[10127-10127x-\frac{2456}{3}x^2+\frac{2456}{3x}\right] 
+64\left[6-6x-\frac{1}{3}x^2+\frac{1}{3x}\right]\zeta(2)
\non\\&&
+128\left[1-x-\frac{4}{3}x^2+\frac{4}{3x}\right]\zeta(3)
\\
P_{12}^{(5)}(x) &=& 
32(1-2x+2x^2)\ln^4(1-x)-64(1-2x+4x^2)\ln(x)\ln^3(1-x)
\non\\&&
-24(1-2x-8x^2)\ln^2(x)\ln^2(1-x)
+\frac{8}{3}(5-10x-16x^2)\ln^3(x)\ln(1-x)
\non\\&&       
+\ln^4(x)\left[\frac{5}{12}-\frac{5}{6}x+\frac{8}{3}x^2\right]
-\frac{64}{3}(2-7x+x^2)\ln^3(1-x)
\non\\&&
-16(8+11x-12x^2)\ln(x)\ln^2(1-x)
+\frac{8}{3}(26+23x-60x^2)\ln^2(x)\ln(1-x)
\non\\&&       
+\ln^3(x)\left[5-\frac{x}{3}+\frac{224}{9}x^2\right]
\non\\&&
-4\ln^2(1-x)\left[\frac{649}{9}-\frac{1478}{9}x
                    +\frac{728}{9}x^2-\frac{32}{3x}\right.
\non\\&&\left.\hspace{7cm}                   
+48\zeta(2)-96\zeta(2)x+96\zeta(2)x^2\right]
\non\\&&
+4\ln(x)\ln(1-x)\left[\frac{1027}{9}-\frac{866}{9}x+\frac{1328}{9}x^2
                      +48\zeta(2)-96\zeta(2)x+192\zeta(2)x^2\right]
\non\\&&
+\ln^2(x)\left[\frac{451}{9}-\frac{1664}{9}x-\frac{1072}{9}x^2
                      +24\zeta(2)-48\zeta(2)x-192\zeta(2)x^2\right]
\non\\&&       
+4\ln(1-x)\left[\frac{6064}{27}-\frac{7721}{27}x
                          +\frac{2056}{27} x^2-\frac{128}{27x}
                          + 32\zeta(2)-112\zeta(2)x+16\zeta(2)x^2\right.
\non\\&&\left.\hspace{7cm}
                          + 64\zeta(3)-128\zeta(3)x+128\zeta(3)x^2\right]
\non\\&&
+\ln(x)\left[\frac{8861}{54}+\frac{24511}{27}x-416x^2-\frac{256}{27x}
                          +128\zeta(2)+176\zeta(2)x-192\zeta(2)x^2\right.
\non\\&&\left.\hspace{7cm}
                          -128\zeta(3)+256\zeta(3)x-512\zeta(3)x^2\right]
\non\\&&
- 384x^2\ln^2(1-x)\Li_2(1-x)+ 288(2x-1)\ln(x)\ln(1-x)\Li_2(1-x)
\non\\&&       
+ 16(1-2x+4x^2)\ln^2(x)\Li_2(1-x)-32(12-3x-10x^2)\ln(1-x)\Li_2(1-x)
\non\\&&       
+\frac{32}{3}(1-5x-12x^2)\ln(x)\Li_2(1-x)+ 768x^2\ln(1-x)\Li_3(1-x)
\non\\&&
-288(2x-1)\ln(x)\Li_3(1-x)-32(11-22x+16x^2)\ln(1-x)\Sf(1-x)
\non\\&&
-16(5-10x-16x^2)\ln(x)\Sf(1-x)
\non\\&&
+8\left[21+34x+\frac{100}{3}x^2
                            +\frac{16}{3x}+48\zeta(2)x^2\right]\Li_2(1-x)
\non\\&&       
+32(12-3x-10x^2)\Li_3(1-x)
-16\left[10+5x-\frac{8}{3}x^2\right]\Sf(1-x)
\non\\&&
-768x^2\Li_4(1-x)+ 32(1-2x+2x^2)\Li_2^2(1-x)
\non\\&&       
- 32(7-14x-6x^2)\SS_{1,3}(1-x)+32(7-14x+8x^2)\SS_{2,2}(1-x)
\non\\&&
+\frac{322519}{648}-\frac{240997}{324}x+\frac{23776}{81}x^2
-\frac{1088}{27x}
+\frac{4}{3}\left[\frac{649}{3}-\frac{1478}{3}x
+\frac{728}{3}x^2-\frac{32}{x}\right]\zeta(2)
\non\\&&
-\frac{128}{3}(2-7x+x^2)\zeta(3)+48(1-2x+2x^2)\zeta(4)
\\
P_{21}^{(5)}(x)&=&
-16\left[2-x-\frac{2}{x}\right]\ln^4(1-x)+32(2-x)\ln(x)\ln^3(1-x)
\non\\&&
+12(2-x)\ln^2(x)\ln^2(1-x)-\frac{20}{3}(2-x)\ln^3(x)\ln(1-x)
- \frac{5}{24}(2-x)\ln^4(x)
\non\\&&
+\frac{32}{3}\left[7-2x-\frac{1}{x}\right]\ln^3(1-x)
-8\left[17-16x+\frac{8}{x}\right]\ln(x)\ln^2(1-x)
\non\\&&
+\frac{4}{3}(41-46x)\ln^2(x)\ln(1-x)+\frac{1}{6}(x-47)\ln^3(x)
\non\\&&
+\ln^2(1-x)\left[\frac{2956}{9}-\frac{1298}{9}x+\frac{64}{3}x^2
-\frac{1456}{9x}+192\zeta(2)-96\zeta(2)x-\frac{192}{x}\zeta(2)\right]
\non\\&&       
-\ln(x)\ln(1-x)\left[\frac{4180}{9}
                 -\frac{542}{9}x+\frac{128}{3}x^2-\frac{256}{9x}
                                     +192\zeta(2)-96\zeta(2)x\right]
\non\\&&
-\ln^2(x)\left[\frac{220}{9}-\frac{1207}{18}x-\frac{32}{3}x^2
                        -\frac{64}{9x}+24\zeta(2)-12\zeta(2)x\right]
\non\\&&
-\ln(1-x)\left[\frac{15442}{27}-\frac{12128}{27}x+\frac{256}{27}x^2
     -\frac{4112}{27x}+224\zeta(2)-64\zeta(2)x-\frac{32}{x}\zeta(2)\right.
\non\\&&\left.\hspace{5cm}
+256\zeta(3)-128\zeta(3)x-\frac{256}{x}\zeta(3)\right]
\non\\&&
+\ln(x)\left[\frac{6373}{54}-\frac{57373}{108}x
                    +\frac{128}{9}x^2+\frac{1504}{27x}
+136\zeta(2)-128\zeta(2)x+\frac{64}{x}\zeta(2)\right.
\non\\&&\left.\hspace{7cm}
+128\zeta(3)-64\zeta(3)x\right]
\non\\&&
+\frac{192}{x}\ln^2(1-x)\Li_2(1-x)+144(2-x)\ln(x)\ln(1-x)\Li_2(1-x)
\non\\&&
- 8(2-x)\ln^2(x)\Li_2(1-x)
- 16\left[3-12x+\frac{10}{x}\right]\ln(1-x)\Li_2(1-x)
\non\\&&       
-\frac{16}{3}\left[5-x+\frac{12}{x}\right]\ln(x)\Li_2(1-x)
-\frac{384}{x}\ln(1-x)\Li_3(1-x)
\non\\&&
-144(2-x)\ln(x)\Li_3(1-x)
+16\left[22-11x+\frac{8}{x}\right]\ln(1-x)\Sf(1-x)
\non\\&&
+ 40(2-x)\ln(x)\Sf(1-x)
-4
\left[34+21x+\frac{16}{3}x^2+\frac{100}{3x}
                                   +\frac{48}{x}\zeta(2)\right]\Li_2(1-x)
\non\\&&
+16\left[3-12x+\frac{10}{x}\right]\Li_3(1-x)
-8\left[11-14x+\frac{52}{3x}\right]\Sf(1-x)
\non\\&&       
+\frac{384}{x}\Li_4(1-x)-16\left[2-x-\frac{2}{x}\right]\Li_2^2(1-x)
\non\\&&
+16\left[14-7x+\frac{2}{x}\right]\SS_{1,3}(1-x)
-16\left[14-7x+\frac{16}{x}\right]\SS_{2,2}(1-x)
\non\\&&
-\frac{240997}{648}+\frac{322519}{1296}x-\frac{544}{27}x^2
+\frac{11888}{81x}
-\left[\frac{2956}{9}-\frac{1298}{9}x
                          +\frac{64}{3}x^2-\frac{1456}{9x}\right]\zeta(2)
\non\\&&       
+\frac{64}{3}\left[7-2x-\frac{1}{x}\right]\zeta(3)
-24\left[2-x-\frac{2}{x}\right]\zeta(4)
\\
\label{eqSP2}
P_{22}^{(5)}(x) &=&
32(1+x)\ln(x)\ln^3(1-x)-24(1+x)\ln^2(x)\ln^2(1-x)
\non\\&&
-\frac{20}{3}(1+x)\ln^3(x)\ln(1-x)+\frac{7}{6}(1+x)\ln^4(x)
\non\\&&
+16\left[1-x-\frac{4}{3}x^2+\frac{4}{3x}\right]\ln^3(1-x)
-\frac{16}{3}(4-5x+12x^2)\ln(x)\ln^2(1-x)
\non\\&&
-\ln^2(x)\ln(1-x)\left[\frac{82}{3}-\frac{74}{3}x+32x^2\right]
+\frac{2}{9}\ln^3(x)\left[35-10x+16x^2\right]
\non\\&&
-\frac{4}{3}\left[77-77x
       -\frac{32}{3}x^2+\frac{32}{3x}\right]\ln^2(1-x)
\non\\&&
-\ln(x)\ln(1-x)\left[\frac{862}{3}+\frac{1478}{3}x+\frac{512}{9}x^2
                     +\frac{256}{9x}+96\zeta(2)+96\zeta(2)x\right]
\non\\&&
+\ln^2(x)\left[\frac{319}{3}+\frac{266}{3}x+\frac{64}{3}x^2+
                                    24\zeta(2)+24\zeta(2)x\right]
\non\\&&
-\ln(1-x)\left[\frac{2401}{9}-\frac{2401}{9}x
                       -\frac{2624}{27}x^2+\frac{2624}{27x}\right.
\non\\&&\left.\hspace{5cm}
+48\zeta(2)-48\zeta(2)x-64\zeta(2)x^2+\frac{64}{x}\zeta(2)\right]
\non\\&&       
+\ln(x)\left[\frac{8680}{27}+\frac{1477}{27}x-\frac{2240}{27}x^2
+\frac{128}{9x}
\right.
\non\\&&\left.\hspace{5cm}
+\frac{64}{3}\zeta(2)-\frac{80}{3}\zeta(2)x-64\zeta(2)x^2
  +64\zeta(3)+64\zeta(3)x \right]
\non\\&&
+96(1+x)\ln^2(1-x)\Li_2(1-x)-44(1+x)\ln^2(x)\Li_2(1-x)
\non\\&& 
+\frac{16}{3}\left[1+x+12x^2+\frac{12}{x}\right]\ln(1-x)\Li_2(1-x)
-76(1-x)\ln(x)\Li_2(1-x)
\non\\&&
-192(1+x)\ln(1-x)\Li_3(1-x)+96(1+x)\ln(1-x)\Sf(1-x)
\non\\&&
-104(1+x)\ln(x)\Sf(1-x)
\non\\&&
-\Li_2(1-x)\left[390+390x+\frac{128}{3}x^2+\frac{128}{3x}
                                     +96\zeta(2)+96\zeta(2)x\right]
\non\\&&
-\frac{16}{3}\left[1+x+12x^2+\frac{12}{x}\right]\Li_3(1-x)
-\frac{4}{3}\left[61-65x-32x^2-\frac{16}{x}\right]\Sf(1-x)
\non\\&&
+192(1+x)\Li_4(1-x)-88(1+x)\SS_{1,3}(1-x)-96(1+x)\SS_{2,2}(1-x)
\non\\&&
+\frac{24043}{36}-\frac{24043}{36}x-\frac{5408}{81}x^2+\frac{5408}{81x}
+\frac{4}{3}\left[77-77x-\frac{32}{3}x^2+\frac{32}{3x}\right]\zeta(2)
\non\\&&
+ 32\left[1-x-\frac{4}{3}x^2+\frac{4}{3x}\right]\zeta(3)
-\frac{32}{243}\del(1-x)
\end{eqnarray}

\noindent
The complexity of the above expressions reaches weight {\sf w = n+p = 4} Nielsen integrals
\begin{eqnarray}
S_{n,p}(x) &=& \frac{(-1)^{n+p-1}}{(n-1)! p!} \int_0^1\frac{dz}{z}
\ln^{n-1}(z) \ln^p(1-zx)~.
\end{eqnarray}
The radiators can be expressed through these functions and polynomials 
thereof as well as rational functions in $x$.

The universal radiator functions  Eq.~(\ref{eqSLO}) can now be attached to the respective 
initial- or final-state radiating light charged fermion or photon lines of any differential
scattering cross section to account for the respective leading order QED corrections. These
radiators generalize the radiators due to  soft-photon exponentiation, valid for $D_{\rm NS}(a(Q^2),x)$  
\cite{Kuraev:1985hb} in the region $x \rightarrow 1$, to general values of $x$ 
and all 
collinear transitions possible. The numerical effect  of the respective radiator depends on
the change of the subsystem kinematics of the differential scattering cross section, which 
usually differs for initial and final state radiation and due to the type of leg encountered. This 
kinematics has to be worked out for the respective process accordingly. Moreover, the 
radiative corrections may strongly depend on the way the kinematic variables 
of the process are 
measured. In case of deeply inelastic scattering investigations of these aspects were 
performed in \cite{DeRujula:1979jj,Blumlein:1989gk,Blumlein:1994ii,Arbuzov:1995id}.
For similar considerations for $e^+e^-$ annihilation see
e.g. Ref.~\cite{Beenakker:1989km}. 

The radiative correction due to the radiator $D_{a_1 a_2}(a(Q^2),x)$ for a 
differential cross section reads
\begin{eqnarray}
\label{RAD1}
\frac{d^l \sigma_{a_1}}{d b_1 ... d b_l} =
\int_0^1 dz D_{a_1 a_2}(a(Q^2),z) \theta(z-z_0^{a_1}) J^{a_1}(\left. b_r\right|_{r=1}^l,z)
\left.\frac{d^l \sigma_{a_2}}{d b_1 ... d b_l} \right|_{b_1 = \hat{b}_1 ...b_l = \hat{b}_1}~.
\end{eqnarray}
Here the $l$ kinematic variables which determine the differential cross section are  $\left. 
b_r\right|_{r=1}^l$. Their rescaled value under changing the momentum 
$p_{a_1} \rightarrow z \cdot p_{a_1}$ resp.  
$p_{a_1} \rightarrow p_{a_1}/z$ for initial or final state radiation, $\left. 
\hat{b}_r\right|_{r=1}^l$, is bounded by $z_0^{a_1}$
for hard radiation. $J^{a_1}(\left. b_r\right|_{r=1}^l,z)$ denotes the corresponding Jacobian
\begin{eqnarray}
J^{a_1}(\left. b_r\right|_{r=1}^l,z)
= \left| \begin{array}{ccc}
         \partial \hat{b}_1/\partial b_1 & \ldots  & \partial \hat{b}_l/\partial b_1\\
         \vdots & & \vdots \\
         \partial \hat{b}_l/\partial b_1 & \ldots  & \partial \hat{b}_l/\partial b_l 
        \end{array} \right| 
\end{eqnarray}
and $d^l \sigma_{a_2}/d b_1 ... d b_l$ is the subsystem differential cross section
for which the line of type $a_1$ is being replaced by a line of type $a_2$.
Eq.~(\ref{RAD1}) may be generalized to the case of more universal 
radiators correspondingly, requiring additional rescaling of variables.

In the above we assumed, that the radiator functions describe collinear radiation along 
outer fermions or photons. However, in various applications also {\sf internal}, nearly
collinear situations may occur. One example is the so-called 3rd \cite{Mo:1968cg} or 
Compton peak \cite{Beenakker:1989km,Blumlein:1989gk,Blumlein:1993ef}
in deeply inelastic scattering. Here a photon being originally virtual  
contributes near to its mass shell in the radiative correction, which gives rise to a 
factorizing collinear process. The universal contributions to these processes can be obtained
from radiators as well.  

Finally we would like to add a remark on small $x$ resummations. In QCD the leading order corrections 
stem from 
\begin{eqnarray} 
\label{eqsx1} 
P_{gg}^{x \rightarrow 0} (x) &=& \M^{-1}[\gamma_L(N,a_s)](x) \\
\label{eqsx2} 
P_{gq}^{x \rightarrow 0} (x) &=& \frac{C_F}{C_A} P_{gg}^{x \rightarrow 0} (x)~. 
\end{eqnarray} 
and the function $\gamma_L(N,a_s)$ obeys 
\begin{eqnarray} 
\gamma_L(N,a_s) = \frac{C_A a_s}{ \pi (N-1)} \left\{1 + 2 \sum_{l=1}^{\infty} 
\zeta_{2l+1} \gamma_L^{2l+1}\right\}~, 
\end{eqnarray} 
with $a_s = \alpha_s/(4\pi)$ the strong coupling constant and 
$\zeta_k$ Riemann's $\zeta$-function. The transition to 
QED, $C_A \rightarrow 0, C_F \rightarrow 1$ trivializes (\ref{eqsx1},\ref{eqsx2}) except for the lowest 
order term in $a(Q^2)$ in $P_{\gamma f}^{x \rightarrow 0} (x)$ which is already 
known from the fixed 
order terms above. Yet all anomalous dimensions do receive $1/x$ terms in higher orders, which 
contribute, as well-known \cite{NLO}, in the abelian limit starting from next-to-leading order~:
\begin{eqnarray} 
\label{eqlx1}
P_{ff}^{(1)}(x)             &\propto& \frac{1}{x}  \frac{40}{9} N_f \\ 
P_{\gamma f}^{(1)}(x)       &\propto& \frac{1}{x}  \frac{40}{9} N_f \\ 
P_{f \gamma}^{(1)}(x)       &\propto& \frac{1}{x}  \frac{40}{9} N_f \\ 
\label{eqlx4}
P_{\gamma \gamma}^{(1)}(x)  &\propto& \frac{1}{x}  \frac{4}{3} N_f~.  
\end{eqnarray} 
However, these terms do not stem from the resummation \cite{Fadin:1975cb} but 
are of different origin. Their pole strength is of $O(\alpha^2/(N-1))$, 
which is 
larger than for the poles resulting from $O(\alpha^2 \ln^2(x))$ terms. 
For unpolarized
QED radiators systematic resummations of the leading small $x$ terms were not
carried out yet. At leading order in $a(Q^2)$ only 
$P_{\gamma f}(x) \propto 1/x$, while 
at next-to-leading order (\ref{eqlx1}--\ref{eqlx4}) all terms contain a singular
contribution. It would be worthwhile to derive  resummations of these terms 
in the future. 

\section{Numerical Results}

\vspace{1mm}
\noindent
The singlet contributions to the universal radiator functions,
summing the leading logarithmic corrections up to ${ O}((\alpha L)^5$),
are shown in Figure~1, as a function of the momentum fraction $x$ 
for different values of $Q=\sqrt{Q^2}$ in the case of $m_f = m_e$. The corrections in case
of other charged fermions have to be rescaled accordingly in $L = 
\ln(Q^2/m_f^2)$.
In case of fractionally charged leptons, $\alpha_0$ has to be replaced by $e_f^2 \alpha_0$.
Here $D_{11}^{PS}$ denotes the pure singlet part of the fermion radiator 
to which the non-singlet contribution has to be added, cf.~\cite{Blumlein:2004bs}. 
The diagonal radiator $D_{11}^{\rm PS}$ vanishes at leading order ${ O}(\alpha L)$, while 
the 
radiator $D_{22}$ contributes at $x=1$ only due to momentum conservation. For $x < 1$ the 
(pure) singlet ($PS$)-diagonal radiators contribute with ${ O}((\alpha L)^2)$ only.  
In this order the $PS$-diagonal terms are identical in the region  $x < 1$ and differ 
by the $\delta(1-x)$-distribution in $D_{22}$. At higher orders  both diagonal elements 
receive different corrections. This explains the relative smallness of the radiators
$D_{11}^{\rm PS}$ and $D_{22}$ compared to $D_{12}$ and $D_{21}$. Yet the diagonal radiators grow
$\propto 1/x$ as $x \rightarrow 0$. This growth is even more pronounced in $D_{21}$, which 
contains a $1/x$ term already at $O(\alpha L)$, while  $D_{12}$ receives those terms at 
$O((\alpha L)^3)$ only and therefore shows a moderately varying profile in $x$.
The QED scaling violations shown in Figure~1 are of moderate size, comparing  scales
from $Q = 10 \GeV$ to $Q = 1 \TeV$ which is due to the smallness of the fine structure 
constant and its weak running.   
At $x=0.1$ the radiators $D_{11}^{\rm PS}$ and $D_{22}$  reach 
about $1\%$ and grow further towards 
smaller values of $x$. $D_{12}$ takes values between $3$ and $8 \%$. $D_{21}$ is largest
and reaches $50\%$ at $x=0.1$. The radiators  $D_{11}^{\rm PS}, D_{21}$ and $D_{22}$  
vanish towards $x \rightarrow 1$, while $D_{12}$ approaches finite values.    

Figure~2 compares the size of the first order contribution to $D_{11}$, 
$D_{12}$ and $D_{21}$ with the respective contributions up to ${ O}((\alpha L)^5)$, which 
can be regarded numerically as the total contribution for 
the values of  $Q$ chosen. For $D_{11}$ the first order contribution is much
smaller than the total contribution in the region of small values $x$ due to the steep 
rise of the pure singlet component. 
The higher order contributions to $D_{12}$ are small at medium values of $x$, 
and amount to a $-5\%$ correction in the small $x$ region and a +10\% 
correction at 
large values of $x$ for $Q=10~\GeV$. The higher order contribution to $D_{21}$ range between 
1\% to 10\%.

Figure~3 shows the impact 
of the 5th order term w.r.t. the first four orders. The effect of 
the 5th order term amounts to ${ O}(10^{-5})$ for $D_{11}$, $D_{12}$ 
and $D_{21}$, while the effect in the case of $D_{22}$ is one order 
of magnitude larger at large value of $x$. In either case, we confirm 
that the singlet radiator to the 5th order has a very high accuracy, 
which is sufficient to represent the universal part of QED corrections,  
relevant in high precision measurements in both high-energy charged lepton 
anti-lepton collision, charged lepton-nucleon collisions and photon collisions at
future electron-positron linear colliders with a possible Giga--$Z$ option, 
electron-photon and photon-photon colliders and future muon colliders. 
These are the reactions to which the corresponding radiators are contributing.
The reactions do also contribute to the precision measurements of QCD scaling 
violations \cite{DeRujula:1979jj,Blumlein:1989gk,QCDVA} in deeply inelastic scattering 
as universal QED corrections. Likewise they are important
for rare initial states at high energy hadron colliders such as LHC and would contribute
in lepton and photon initiated processes there, such as single leptoquark production
\cite{LQ}.

The radiator functions calculated above are made available in form of a 
{\tt FORTRAN}-program which can be obtained form the authors upon request.  

\section{Conclusions}

\vspace{1mm}
\noindent
The collinear logarithms  in QED can be resummed due to the 
renormalization group equations for mass 
factorization. Unlike the case in QCD the collinear logarithms are finite due to the 
possibility to define the coupling constant asymptotically, i.e. in the limit of vanishing scales.
The associated logarithms are well defined since photons and leptons
are non-confined and may be regarded as  stable 
or long--lived states. 
The leading order corrections $O((\alpha L)^k)$ are universal. The respective 
radiators resum the radiative corrections which only depend on the type of particle transition $i 
\rightarrow j$. As shown, sufficient numerical stability of $O(10^{-4 ... 
-5})$ is reached evaluating the
radiators to $O((\alpha L)^5)$ for scales as large as $Q \lsim 1 \TeV$. 
The radiators are presented in $x$--space and can be applied directly to the respective multiply 
differential scattering cross sections to describe the universal contribution due to initial and final 
state radiation off the different outer legs contributing to the respective scattering process involving 
charged fermions and photons. In the small $x$ region 
the leading  order radiators receive contributions $\propto 1/x$, with an onset in different orders in 
$\alpha L$, which leads to larger corrections in this kinematic region. A systematic resummation of the
particular small $x$ contributions, unlike the case for the non--singlet and polarized singlet 
corrections, is not known yet.  
The radiators derived can easily be adopted for experimental analysis and simulation programs.

\newpage
\section{Figures}

\vspace{1cm}
\noindent
\begin{figure}[ht]
\begin{center}
\includegraphics[height=12cm]{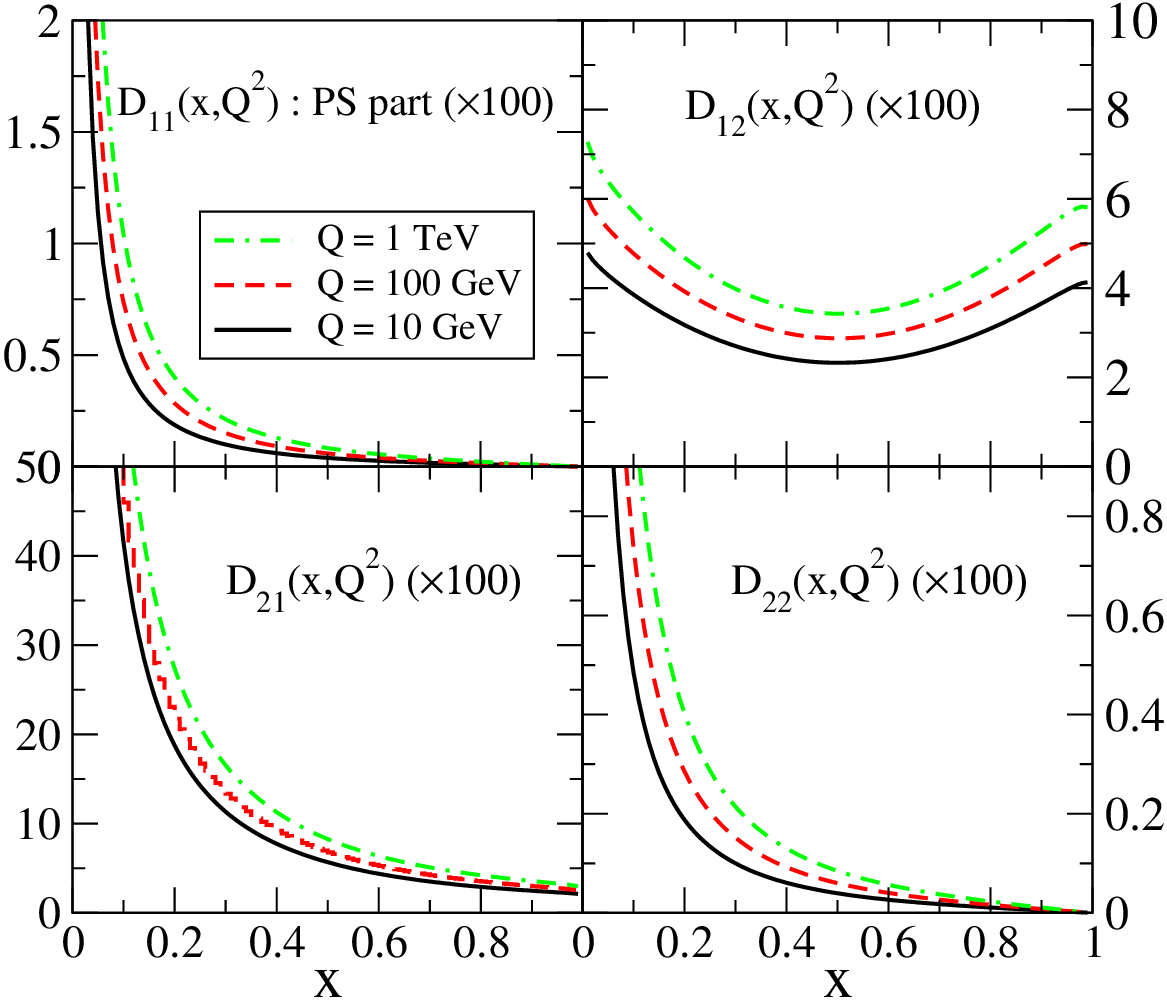}
\end{center}
\caption{The singlet radiators $D_{ij}$ as a function of $x$ 
and $Q$ in $\%$. $D_{11}^{PS}$ denotes the pure singlet part of $D_{11}$.}
\end{figure}

\newpage

\vspace{3cm}
\noindent
\begin{figure}[ht]
\begin{center}
\includegraphics[height=12.cm]{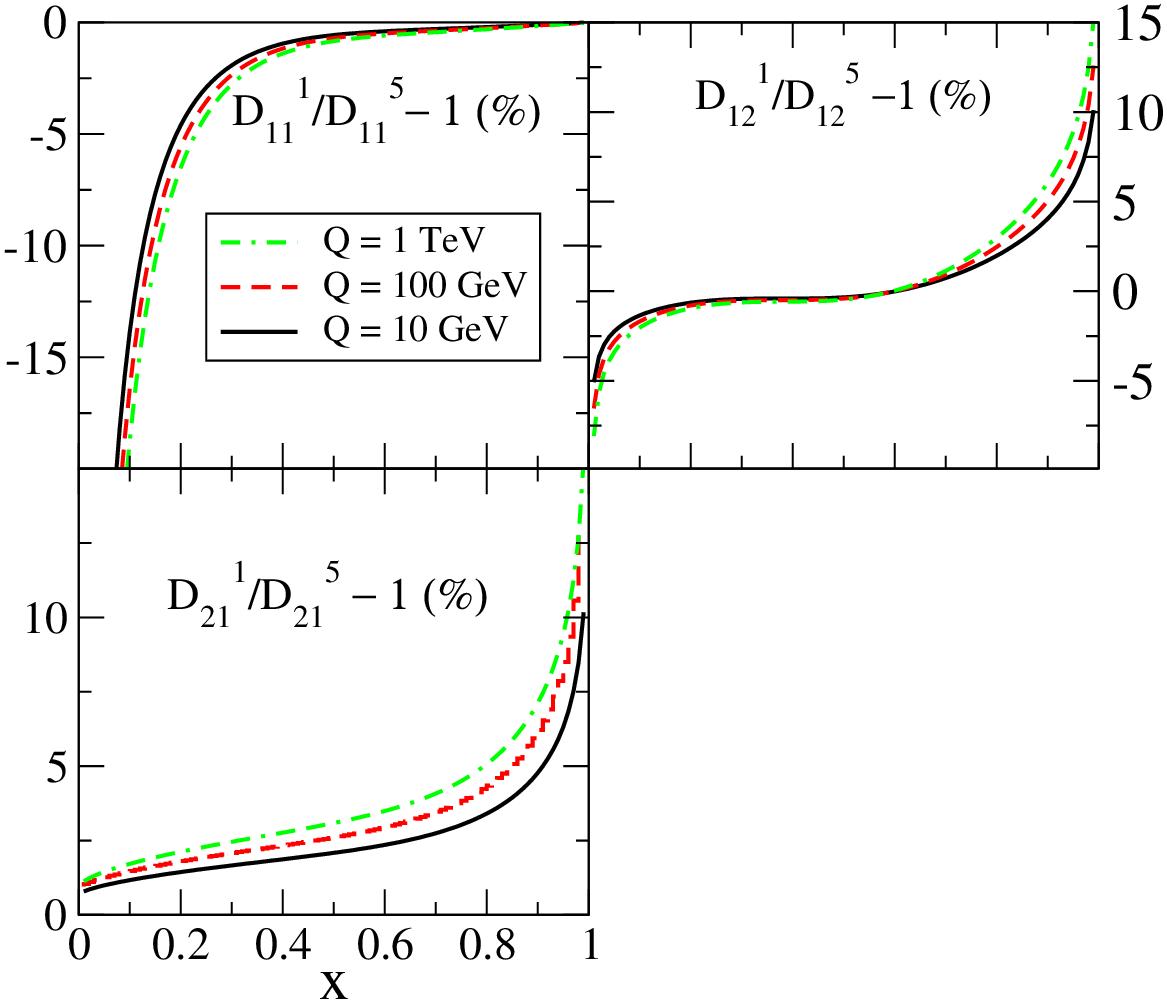}
\end{center}
\caption{Relative contribution of the first order singlet radiators 
$D^1_{ij}$ in all terms to ${ O}((\alpha L)^5)$.
Here $D_{11}$ denotes the sum of the non-singlet contributions $D_{\rm NS}$, 
with soft exponentiation beyond  ${ O}((\alpha L)^5)$, and the pure singlet contribution $D_{11}^{\rm 
PS}$.}
\end{figure}

\newpage

\vspace{3cm}
\noindent
\begin{figure}[h]
\begin{center}
\includegraphics[height=12cm]{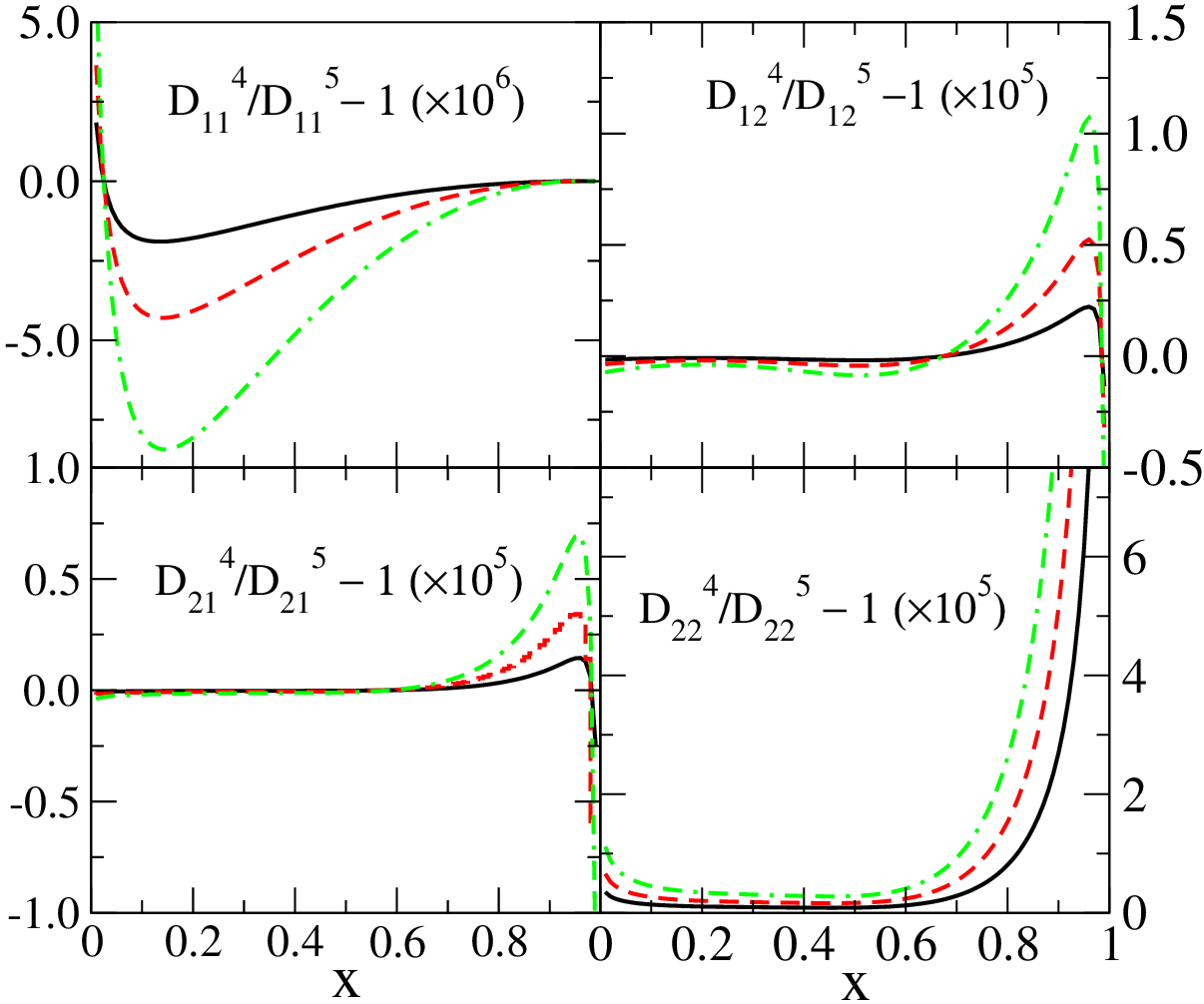}
\end{center}
\caption{Relative contribution of the singlet radiators $D_{ij}$ 
up to the 4th order in $\alpha L$ if compared to all terms to the 5th 
order as a function of $x$ and $Q$, with $D_{11} = D_{11}^{\rm NS+PS}$.}
\end{figure}

\newpage
\section{Appendix A:~Mellin Convolutions}

\vspace{1mm}
\noindent
In this appendix we list the convolutions of functions up to weight~5 required in
the present calculation in addition to those given in Ref.~\cite{Blumlein:2004bs,Blumlein:2000wh}.
Some of the integrals require to use Mellin transforms and algebraic relations between the
finite harmonic sums \cite{Blumlein:1999if,Blumlein:2003gb}.
They were calculated recursively in explicit form and may be
of general interest for other higher order calculations in QED and QCD. 
\normalsize
\begin{eqnarray}
1 \otimes \frac{1}{x} &=& \frac{1}{x}-1 \\
x \otimes \frac{1}{x} &=& \frac{1}{2}\left(\frac{1}{x}-x\right) \\
x^2 \otimes \frac{1}{x} &=& \frac{1}{3}\left(\frac{1}{x}-x^2\right)\\
\frac{1}{x} \otimes \frac{1}{x} &=&  -\frac{1}{x} \ln(x) \\
\left(\frac{1}{1-x}\right)_+ \otimes \frac{1}{x} &=& \frac{1}{x} \ln(1-x) \\
\left(\frac{\ln(1-x)}{1-x}\right)_+ \otimes \frac{1}{x} &=&
\frac{1}{2x}\ln^2(1-x)\\
\ln(1-x) \otimes \frac{1}{x} &=&  \left(\frac{1}{x}-1\right)[\ln(1-x)-1]
\\
x\ln(1-x) \otimes \frac{1}{x} &=& \frac{1}{2}\left(\frac{1}{x}-x\right)
\ln(1-x)+\frac{x}{4}+\frac{1}{2}-\frac{3}{4x} 
\\
\ln(x) \otimes \frac{1}{x} &=& -\ln(x)+1-\frac{1}{x} 
\\
x\ln(x) \otimes \frac{1}{x} &=& -\frac{x}{2}\ln(x)
+\frac{1}{4}\left(x-\frac{1}{x}\right)
\\
\frac{\ln(x)}{1-x} \otimes \frac{1}{x} &=& -\frac{1}{x}\Li_2(1-x) 
\\
\left(\frac{1}{1-x}\right)_+ \otimes x^2 \ln(1 - x) &=& 
x^2\Biggl[\ln^2(1-x)-\ln(x)\ln(1-x)-\zeta_2\Biggr]\non\\
&&+\frac{1}{2}\left(1+2x-3x^2\right)\ln(1-x)
+\frac{5}{4}x^2-\frac{x}{2}-\frac{3}{4}
\\
\left(\frac{1}{1-x}\right)_+ \otimes x^2 \ln(x) &=& 
x^2\Biggl[-\frac{1}{2}\ln^2(x)+\Li_2(1-x)+\ln(x)\ln(1-x)\Biggr]\\
&&-\frac{3}{2}x^2\ln(x)-\frac{1}{4}-x+\frac{5}{4}x^2
\non\\
1 \otimes x^2 \ln(1 - x) &=& \frac{1}{2}(1-x^2)\ln(1-x)+\frac{x^2}{4}
+\frac{x}{2}-\frac{3}{4}
\\
1 \otimes x^2 \ln(x) &=& -\frac{x^2}{2}\ln(x)-\frac{1}{4}(1-x^2)
\\
x \otimes x^2 \ln(1-x) &=& x(1-x)[\ln(1-x)-1]
\end{eqnarray}\begin{eqnarray}
x \otimes x^2 \ln(x) &=& -x[1-x+x\ln(x)]
\\
1 \otimes x^2 &=& \frac{1}{2}(1-x^2)
\\
x \otimes x^2 &=& x(1-x)
\\
x^2 \otimes x^2 &=& -x^2\ln(x)
\\
x^2 \otimes \ln(x) &=& \frac{1}{2}\ln(x)+\frac{1}{4}(1-x^2)
\\
x^2 \otimes x\ln(x) &=& x[1-x+\ln(x)]
\\
\frac{1}{x} \ln(1-x) \otimes \frac{1}{x} 
&=& \frac{1}{x}[\Li_2(x)-\zeta_2] 
\\
x^2 \ln(1-x) \otimes \frac{1}{x} 
&=& \frac{1}{3}\left(\frac{1}{x}-x^2\right)\ln(1-x) 
- \frac{11}{18x} +\frac{1}{3} + \frac{x}{6} + \frac{x^2}{9} 
\\
x^2 \ln(x) \otimes \frac{1}{x} 
&=& -\frac{x^2}{3} \ln(x) + \frac{1}{9} (x^2-\frac{1}{x}) 
\\
\left(\frac{\ln^2(1-x)}{1-x}\right)_+  \otimes \frac{1}{x}
&=& \frac{1}{3x} \ln^3(1-x)
\\
\ln^2(1-x)  \otimes \frac{1}{x}
&=& \left(\frac{1}{x}-1\right)[ \ln^2(1-x)-2 \ln(1-x) +2 ]
\\
x \ln^2(1-x)  \otimes \frac{1}{x}
&=&  \frac{1}{2}\left(\frac{1}{x}-x\right) \ln^2(1-x) 
+\left(\frac{x}{2}+1-\frac{3}{2x}\right) \ln(1-x) 
\nonumber \\ & &
+\frac{7}{4x}-\frac{3}{2}-\frac{x}{4}
\\
x^2 \ln^2(1-x)  \otimes \frac{1}{x}
&=& \frac{1}{3} \left(\frac{1}{x}-x^2\right) \ln^2(1-x)
+\left(\frac{2}{9}x^2+\frac{x}{3}
+\frac{2}{3}-\frac{11}{9x}\right)\ln(1-x)
\\&&
+\frac{85}{54x}-\frac{11}{9}-\frac{5}{18}x-\frac{2}{27}x^2
\non\\
\frac{\ln(x) \ln(1-x)}{1-x} \otimes \frac{1}{x}
&=& \frac{1}{x} [\Li_3(1-x) - \ln(1-x)\Li_2(1-x)]
\\
\ln(x) \ln(1-x) \otimes \frac{1}{x}
&=&-\frac{1}{x}\Li_2(1-x)-\ln(x)\ln(1-x)
+\left(1-\frac{1}{x}\right)\ln(1-x) \\
&&+\ln(x)-2\left(1-\frac{1}{x}\right)
\non\\
x \ln(x) \ln(1-x)  \otimes \frac{1}{x}
&=& -\frac{1}{2x}\Li_2(1-x)-\frac{x}{2}\ln(x)\ln(1-x)
+\frac{1}{4}\left(x-\frac{1}{x}\right)\ln(1-x)\\
&&+\frac{1}{4}(2+x)\ln(x)+\frac{1}{x}-\frac{3}{4}-\frac{x}{4}
\non\\
x^2 \ln(x) \ln(1-x)  \otimes \frac{1}{x}
&=&-\frac{1}{3x}\Li_2(1-x)-\frac{x^2}{3}\ln(x)\ln(1-x)
+\frac{1}{9}\left(x^2-\frac{1}{x}\right)\ln(1-x)\non\\
&&+\frac{1}{18}(6+3x+2x^2)\ln(x)
+\frac{71}{108x}-\frac{4}{9}
-\frac{5}{36}x-\frac{2}{27}x^2
\\
\frac{\ln^2(x)}{1-x}  \otimes \frac{1}{x}
&=& \frac{2}{x}  \SS_{1,2}(1-x)
\\
\ln^2(x)  \otimes \frac{1}{x}
&=& \frac{2}{x} - 2[1-\ln(x)] -\ln^2(x)
\\
x \ln^2(x)  \otimes \frac{1}{x}
&=& \frac{1}{4} \left[\frac{1}{x}  - x \right] + \frac{x}{2} 
\left[\ln(x) - \ln^2(x)\right]
\\
x^2 \ln^2(x)  \otimes \frac{1}{x}
&=& \frac{2}{27} \left[\frac{1}{x} -x^2\right] + x^2 \left[\frac{2}{9}
\ln(x) - \frac{1}{3} \ln^2(x)\right]
\\
\Li_2(1-x) \otimes \frac{1}{x}
&=& \left(\frac{1}{x} - 1 \right) \left[\Li_2(1-x)-1\right] - \ln(x) 
\\
x \Li_2(1-x) \otimes \frac{1}{x}
&=&
\frac{1}{2} \left(\frac{1}{x} -x \right) \Li_2(1-x) - \frac{1}{4} 
(2+x) \ln(x) - \frac{5}{8x} + \frac{1}{2} + \frac{1}{8} x
\\
x^2 \Li_2(1-x) \otimes \frac{1}{x}
&=& \frac{1}{3} \left (\frac{1}{x} -x^2\right) \Li_2(1-x)
- \left(\frac{1}{3} + \frac{x}{6} + \frac{x^2}{9} \right) \ln(x)
\nonumber\\ & &
- \frac{49}{108x}+\frac{1}{3} + \frac{x}{12} +\frac{x^2}{27}  
\\
\left(\frac{1}{1-x}\right)_+ \otimes x^2\ln(x)
&=& x^2 \left[ \zeta_2-\Li_2(x)-\frac{1}{2}\ln^2(x)
-\frac{3}{2}\ln(x) \right]-\frac{1}{4}-x+\frac{5}{4}x^2 \\
x^2 \otimes \left(\frac{\ln^2(1-x)}{1-x}\right)_+
&=& \frac{x^2}{3} \ln^3(1-x) + \left(\frac{1}{2} + x - \frac{3}{2} x^2
\right) \ln^2(1-x) 
\nonumber\\ & &-x(1-x)  
\ln(1-x) -x^2 \ln(x) \nonumber\\ & & 
+ 2x^2\left[\zeta_3-\SS_{1,2}(x) \right]
+3x^2\left[\zeta_2 - \Li_2(x)\right] 
\\
x^2 \otimes \ln^2(1-x)
&=& \frac{1}{2}\left(1-x^2\right) \ln^2(1-x) - x(1-x) \ln(1-x)\nonumber\\ 
& & + x^2 \left[\zeta_2 - \Li_2(x) - \ln(x) \right]
\\
x^2 \otimes x \ln^2(1-x)
&=& x(1-x) \ln^2(1-x) - 2 x^2 \left[\Li_2(x) - \zeta_2\right]
\\
x^2 \otimes \frac{\ln(x) \ln(1-x)}{1-x}
&=& 
x^2[2 \zeta_3-\Li_3(x)-\Sf(x)]+\frac{3}{2}x^2\Li_2(1-x)
\nonumber\\ & &      
+\frac{3}{4}x^2\ln^2(x)
+\frac{1}{4}(1+4x-5x^2)\ln(1-x)
\nonumber\\ & &
   +\ln(x)\Biggl[\frac{x^2}{2}\ln^2(1-x)+x^2\Li_2(x)
+\left(x+\frac{1}{2}\right)\ln(1-x)
\nonumber\\ & &            
+\frac{5}{4}x^2-\frac{x}{2}\Biggr]-\frac{3}{4}x(1-x)
\\
x^2 \otimes \ln(x) \ln(1-x)
&=& 
\frac{x^2}{2}\Li_2(1-x)+\frac{1}{4}(1-x^2)\ln(1-x)
    +\frac{x^2}{4}\ln^2(x)
\nonumber\\ & &+\frac{1}{2}\left[\ln(1-x)-x+\frac{x^2}{2}\right]\ln(x)
    -\frac{3}{4}x(1-x)
\\
x^2 \otimes x \ln(x) \ln(1-x)
&=& x(1-x)[1+\ln(x)]\ln(1-x) + x^2\left[\zeta_2 - 
\Li_2(x)\right] 
\nonumber\\ & &+ x^2 \ln(x) \left(1+ \frac{1}{2} \ln(x)\right)
\\
x^2 \otimes \frac{\ln^2(x)}{1-x} 
&=& 2 x^2 \SS_{1,2}(1-x) -\frac{x^2}{3} \ln^3(x)
+ \left(\frac{1}{2} +x \right) \ln^2(x) \nonumber\\
& &+\left(\frac{1}{2} +2 x\right) \ln(x) 
+\frac{1}{4} +2x - \frac{9}{4} x^2 
\\
x^2 \otimes \ln^2(x) 
&=& \frac{1}{4}\left(1-x^2\right) +\frac{1}{2}\left[\ln(x)+\ln^2(x)\right]
\\
x^2 \otimes x \ln^2(x)
&=& 2x(1-x) +x \left[\ln^2(x) +2 \ln(x) \right]
\\
x^2 \otimes \Li_2(1-x)
&=& \frac{1}{2} \left(1-x^2\right) \Li_2(1-x)  +\frac{1}{2} x(1-x) 
+\frac{x}{2} \left[\ln(x) - \frac{x}{2} \ln^2(x)\right]
\\
x^2 \otimes x \Li_2(1-x)
&=& x(1-x) \Li_2(1-x) - \frac{x^2}{2} \ln^2(x)
\\
\left(\frac{\ln^3(1-x)}{1-x}\right)_+  \otimes \frac{1}{x}
&=& \frac{1}{4x} \ln^4(1-x)
\\
\frac{\ln(x)\ln^2(1-x)}{1-x}  \otimes \frac{1}{x}
&=& \frac{1}{x}\left\{\frac{1}{3} \ln(x) \ln^3(1-x) + 2 \left[\SS_{1,3}(x) 
- \zeta_4\right] \right\}
\\
\frac{\ln^2(x)\ln(1-x)}{1-x}  \otimes \frac{1}{x}
&=&  \frac{1}{x}  \left\{\frac{1}{2} \ln^2(x) \ln^2(1-x) + 2 \left[
\SS_{2,2}(x) -
\ln(x) \SS_{1,2}(x) - \frac{\zeta_4}{4}\right]
\right\}
\\
\frac{\ln^3(x)}{1-x}  \otimes \frac{1}{x}
&=& -\frac{6}{x} \SS_{1,3}(1-x)  
\\
\frac{\Sf(1-x)}{1-x}  \otimes \frac{1}{x}
&=& \frac{1}{x} \SS_{2,2}(1-x)
\\
\frac{\Li_3(1-x)}{1-x}  \otimes \frac{1}{x}
&=& \frac{1}{x} \Li_4(1-x)
\\
\frac{\ln(x)\Li_2(1-x)}{1-x}  \otimes \frac{1}{x} 
&=& -\frac{1}{2x} \Li^2_2(1-x)
\\
\frac{\Li_3(x)-\zeta_3}{1-x}  \otimes \frac{1}{x} 
&=& \frac{1}{x} \Biggl\{\frac{1}{2} \left[\Li_2^2(x) -\zeta_2^2\right] 
+ \ln(1-x) \left[\Li_3(x)-\zeta_3\right] \Biggr\}
\\
\Sf(x)  \otimes \frac{1}{x}
&=&  \frac{\zeta_3}{x} - \SS_{1,2}(x) - \frac{1}{x}(1-x) \left[\frac{1}{2}
\ln^2(1-x) - \ln(1-x) +1\right]
\\ 
x \Sf(x)  \otimes \frac{1}{x}
&=& \frac{1}{2x} \left[\zeta_3 - x^2 \SS_{1,2}(x)\right]
-\frac{1}{8x} \left(1-x^2\right) \ln^2(1-x) \nonumber\\ & &
+\frac{1}{8x} \left(3-2x-x^2\right) \ln(1-x)
- \frac{1}{x}\left(\frac{7}{16}
-\frac{3}{8}x -\frac{1}{16} x^2\right)
\\
\Li_3(1-x)  \otimes \frac{1}{x} 
&=& \left(\frac{1}{x} -1 \right) \left[\Li_3(1-x) - \Li_2(1-x) +1 \right]
+ \ln(x)
\\
x \Li_3(1-x)  \otimes \frac{1}{x}
&=& \frac{1}{2x} (1-x^2) \Li_3(1-x) - \left(\frac{3}{4x} -\frac{1}{2} - 
\frac{1}{4} x\right) \Li_2(1-x) \nonumber\\
& & + \left(\frac{3}{4} + \frac{x}{8} \right) \ln(x) + \frac{13}{16 x}
-\frac{3}{4} - \frac{x}{16}
\\
\ln(1-x)\Li_2(1-x)  \otimes \frac{1}{x}
&=& -\frac{1}{x}(1-x) \left[1 -\ln(1-x)\right] \Li_2(1-x)
\\ & &
-\frac{1}{x} (1-x) \left[1 - \ln(x)\right] \ln(1-x)
\nonumber\\ & &
+2 \ln(x) + \frac{1}{x}\left[\Li_2(x) - \zeta_2\right]
+ \frac{3}{x}(1-x)
\non\\
x \ln(1-x)\Li_2(1-x)  \otimes \frac{1}{x}
&=& 
\frac{1-x}{x} \Biggl\{\left[\ln(1-x)-1\right] - \frac{1-x}{2}
\left[\ln(1-x) - \frac{1}{2} \right] \Biggr\}\Li_2(1-x)
\nonumber\\ & &+
\frac{1}{x} \left\{\left[\frac{3}{4} - \frac{x}{2} - \frac{x^2}{4}
\right] \ln(1-x) + \frac{3}{2} x + \frac{x^2}{4} \right\} \ln(x)
\\ & &
- \frac{1}{8x} \left[5 - 4x -x^2\right] \ln(1-x) 
+ \frac{3}{4x} \left[\Li_2(x) - \zeta_2\right]
\non\\&&
+ \frac{37}{16x} - \frac{17}{8} - \frac{3}{16}x 
\non\\
\ln(x)\Li_2(1-x)  \otimes \frac{1}{x}
&=&
-\frac{1}{x} \Biggl\{  2 \SS_{1,2}(1-x) + x \ln^2(x)
\nonumber\\ & &
+ \left[1-x+x \ln(x)\right] \left[\Li_2(1-x) - 3\right] \Biggr\}
\\
x \ln(x)\Li_2(1-x) \otimes \frac{1}{x}
&=& -\frac{1}{x} \SS_{1,2}(1-x) - \frac{1}{4x} \left[1 - x^2 + 2 x^2
\ln(x) \right]
\Li_2(1-x) 
\nonumber\\ & &
- \frac{1}{4} \left(2+x\right) \ln^2(x) 
+\frac{1}{8} \left(10 + 3 x\right) \ln(x)
\nonumber\\ & &
+ \left(\frac{23}{16x} - \frac{5}{4} - \frac{3}{16} x\right) 
\\
\Sf(1-x) \otimes \frac{1}{x}
&=& \left(\frac{1}{x} -1\right) \left[\SS_{1,2}(1-x) -1\right]
+ \frac{1}{2} \ln^2(x) - \ln(x) 
\\
x \Sf(1-x) \otimes \frac{1}{x}
&=& \frac{1}{2}\left(\frac{1}{x}-x\right) \SS_{1,2}(1-x)
+\frac{1}{8} \left(2+x\right) \ln^2(x) \nonumber\\
& &-\frac{1}{8}
\left(4+ x\right) \ln(x)
-\frac{9}{16x} + \frac{1}{2} +
\frac{x}{16}
\\
\Li_3(x)  \otimes \frac{1}{x}
&=&
-\left(\frac{1}{x} - 1 \right) \left[\ln(1-x) -1 \right] - \frac{1}{x}
\left(\zeta_2 - \zeta_3\right) + \Li_2(x) - \Li_3(x) 
\\
x \Li_3(x)  \otimes \frac{1}{x} &=& 
\frac{1}{x} \left(\frac{3}{16} -\frac{1}{4} \zeta_2 + \frac{1}{2} 
\zeta_3\right) - \frac{1}{8} \left(1+ \frac{x}{2}\right) \nonumber\\
& & -\frac{1}{8} \left( \frac{1}{x} -x\right)\ln(1-x) + x \left[
\frac{1}{4} \Li_2(x) - \frac{1}{2} \Li_3(x)\right]
\\
\ln(x)\Li_2(x)  \otimes \frac{1}{x} &=& \frac{1}{x} \Biggl\{
x\left[1-\ln(x)\right]  \Li_2(x) - \zeta_2 - \Li_2(1-x) \nonumber\\ & &
+3(1-x) - \left[2(1-x) +x \ln(x) \right] \ln(1-x)
+x\ln(x) \Biggr\}
\\
x \ln(x)\Li_2(x)  \otimes \frac{1}{x} &=& 
\frac{1}{4x}\left\{x^2\left[1-2\ln(x)\right]\Li_2(x) - \zeta_2\right\}
\nonumber\\ & & - \frac{1}{4x} \left[1-x^2+x^2\ln(x)\right] \ln(1-x)
\nonumber\\ & & + \frac{11}{16x} - \frac{1}{2} - \frac{3}{16}
x + \frac{1}{8} (x+2) \ln(x) - \frac{1}{4x} \Li_2(1-x)
\\
\ln^3(1-x) \otimes \frac{1}{x} &=& 
\left(\frac{1}{x} -1 \right)
\left[\ln^3(1-x) -3 \ln^2(1-x) + 6 \ln(1-x) -6 \right] 
\\
x \ln^3(1-x)  \otimes \frac{1}{x} &=&
\frac{1}{2}\left(\frac{1}{x} -x\right) \ln^3(1-x) + \left(-\frac{9}{4x}
+\frac{3}{2} +\frac{3}{4} x\right) \ln^2(1-x) 
\\ & &
+ \left(\frac{21}{4 x} - \frac{9}{2} - \frac{3}{4} x\right) \ln(1-x)
- \frac{45}{8x} + \frac{21}{4} +\frac{3}{8} x
\non\\ 
\ln(x)\ln^2(1-x)  \otimes \frac{1}{x} &=& 
\frac{1}{x}(1-x) \left[\ln(x)-1\right]\ln^2(1-x) 
\\&&
+\frac{4}{x} \left[1-x+\frac{x}{2}\ln(x)\right] \ln(1-x)
\non\\ & & 
- 2 \ln(x)- \frac{2}{x} \left[\Sa_{1,2}(x) - \zeta_3\right] 
+ \frac{2}{x}\Li_2(1-x)- \frac{6}{x}(1-x)
\non\\
x \ln(x)\ln^2(1-x)  \otimes \frac{1}{x} &=& 
\frac{1}{x} \left\{\zeta_3-\SS_{1,2}(x) 
- \frac{3}{2} \left[\Li_2(x) - \zeta_2 \right] \right\}
\\ & & 
-\frac{1}{4x} \left(1-x^2\right)\left[1- 2 \ln(x) \right]\ln^2(1-x) 
\nonumber\\ & &
+\left\{\frac{2}{x} - \frac{3}{2} - \frac{x}{2} + \frac{1}{x}
\left[-\frac{3}{2} + x + \frac{x^2}{2} \right] \ln(x) \right\}\ln(1-x)
\nonumber\\ & &
- \frac{1}{4} (x+6) \ln(x)
- \frac{31}{8x}+ \frac{7}{2} + \frac{3}{8} x
\non\\
\ln^2(x)\ln(1-x)  \otimes \frac{1}{x} &=& 
\frac{2}{x} \left[\SS_{1,2}(1-x)+\Li_2(1-x) +\ln(1-x)-3(1-x)\right]
\\& & 
+ \ln(x)\left[\ln(x)-4\right] - \left[\ln^2(x) -2 \ln(x) +2 \right]
\ln(1-x)
\non\\
x \ln^2(x)\ln(1-x) \otimes \frac{1}{x} &=&
\frac{1}{x} \left[\SS_{1,2}(1-x) + \frac{1}{2}\Li_2(1-x)\right]
\\ & &
+ \frac{1}{4x}\left(1-x^2\right) \ln(1-x) 
+\frac{1}{2} \left[1 + \frac{x}{2} -x \ln(1-x)\right] \ln^2(x)
\nonumber\\ & & 
- \frac{1}{2} \left[3 + x - x \ln(1-x) \right] \ln(x)
-\frac{17}{8x}+\frac{7}{4} + \frac{3}{8} x
\non\\
\ln^3(x)  \otimes \frac{1}{x} &=& -6\left(\frac{1}{x} -1 \right) -
\ln^3(x) +3 \ln^2(x) -6 \ln(x)  \\
x \ln^3(x) \otimes \frac{1}{x} &=& - \frac{3}{8} \left(\frac{1}{x}
-x \right) -\frac{x}{2} \left[ \ln^3(x) - \frac{3}{2} \ln^2(x)
+\frac{3}{2} \ln(x) \right]
\\
\frac{1}{x} \Li_2(1-x) \otimes \frac{1}{x} &=& 
-\frac{1}{x} \left[2 \SS_{1,2}(1-x) + \ln(x) \Li_2(1-x)\right]
\\ 
\frac{1}{x} \Li_2(1-x) \otimes 1 &=&
\left(\frac{1}{x} - 1 \right) \Li_2(1-x) - \frac{1}{2} \ln^2(x)
\\
\frac{1}{x} \Li_2(1-x) \otimes x &=&
\frac{1}{2}(1-x) + \frac{1}{2}\left(\frac{1}{x} - x\right) \Li_2(1-x)
+\frac{1}{2}\ln(x)\left[1-\frac{1}{2} x \ln(x)\right]
\\
\frac{1}{x} \Li_2(1-x) \otimes x^2 &=&
\frac{1}{3}\left(\frac{1}{x} - x^2\right) \Li_2(1-x) 
\nonumber\\ & &+\frac{1}{6} 
\ln(x)\left[1+2x-x^2\ln(x)\right] + \frac{1}{12}\left[1+4x-5x^2\right]
\\
1 \otimes x^2\ln^2(1-x) 
&=&\frac{1}{2}(1-x^2)\ln^2(1-x)+\frac{1}{2}(-3+2x+x^2)\ln(1-x)
\\&&
+\frac{7}{4}-\frac{3}{2}x-\frac{x^2}{4}
\non\\
1 \otimes x^2\ln(x)\ln(1-x)
&=&\frac{1}{2}[\Li_2(x)-\zeta_2]+\frac{1}{2}(1-x^2)\ln(x)\ln(1-x)
\\&&
-\frac{1}{4}(1-x^2)\ln(1-x)+\frac{x}{4}(2+x)\ln(x)
+1-\frac{3}{4}x-\frac{x^2}{4}
\non\\
1 \otimes x^2\ln^2(x)
&=& \frac{x^2}{2}[\ln(x)-\ln^2(x)]+\frac{1}{4}(1-x^2) 
\\
1 \otimes \frac{1}{x}\ln(1-x) 
&=& \left(\frac{1}{x}-1\right)\ln(1-x)+\ln(x)  
\\
1 \otimes x^2\Li_2(1-x)
&=& \frac{1}{2}(1-x^2)\Li_2(1-x)-\frac{x}{4}(2+x)\ln(x)
-\frac{5}{8}+\frac{x}{2}+\frac{x^2}{8} 
\\
x \otimes x^2\ln^2(1-x)
&=& x(1-x)[\ln^2(1-x)-2\ln(1-x)+2]
\\
x \otimes x^2\ln(x)\ln(1-x)
&=& -x\Li_2(1-x)+x(1-x)[2-\ln(1-x)]
\\&&
+x^2\ln(x)[1-\ln(1-x)]
\non\\
x \otimes x^2\ln^2(x) 
&=& -x^2\ln^2(x)+2x[x\ln(x)-x+1]
\\
x \otimes \frac{1}{x}\ln(1-x)
&=& \frac{1}{2}\left(\frac{1}{x}-x\right)\ln(1-x)+\frac{x}{2}\ln(x)
-\frac{1}{2}(1-x)
\\
x^2 \otimes \frac{1}{(1-x)_+}
&=& x^2[\ln(1-x)-\ln(x)]+\frac{1}{2}+x-\frac{3}{2}x^2
\\
x^2 \otimes \ln(1-x)
&=& \frac{1}{2}(1-x^2)\ln(1-x)+\frac{x}{2}[x\ln(x)+x-1]
\\
x^2 \otimes x\ln(1-x)
&=& x(1-x)\ln(1-x)+x^2\ln(x)
\\
x^2 \otimes \left(\frac{\ln(1-x)}{1-x}\right)_+
&=&-x^2\Li_2(1-x)+\frac{x^2}{2}\ln^2(1-x)-x^2\ln(x)\ln(1-x)
\\&&
+\frac{3}{2}x^2\ln(x)+\left(\frac{1}{2}+x-\frac{3}{2}x^2\right)\ln(1-x)
-\frac{1}{2}x(1-x)
\non\\
x^2 \otimes \frac{\ln(x)}{1-x}
&=& -x^2\Li_2(1-x)-\frac{x^2}{2}\ln^2(x)
+\left(x+\frac{1}{2}\right)\ln(x)
\\&&
+\frac{1}{4}(1+4x-5x^2)
\non
\\
x^2 \otimes \frac{1}{x}\ln(1-x)
&=& \frac{1}{3}\left(\frac{1}{x}-x^2\right)\ln(1-x)+\frac{x^2}{3}\ln(x)
-\frac{1}{6}-\frac{x}{3}+\frac{x^2}{2}  
\\
x^2 \otimes x^2\ln(1-x)
&=& x^2[\Li_2(x)-\zeta_2]  
\\
x^2 \otimes x^2\ln(x)
&=& -\frac{x^2}{2}\ln^2(x)  
\\
x \otimes x^2\Li_2(1-x)
&=& x(1-x)[\Li_2(1-x)-1]-x^2\ln(x)  
\\
x^2 \otimes \left(\frac{\ln^3(1-x)}{1-x}\right)_+ 
&=& 3 x^2 \left\{2 \left[\Sa_{1,3}(x) - \zeta_4\right]
               + 3 \left[\Sa_{1,2}(x) - \zeta_3\right]
               +   \left[\Li_2(x)     - \zeta_2\right] \right\}
\\ & &
+\frac{x^2}{4}\ln^4(1-x) + \frac{1}{2}\left(1+2x-3x^2\right) \ln^3(1-x)
\nonumber\\ & & -\frac{3}{2} x(1-x) \ln^2(1-x)   
\non\\
x^2 \otimes \frac{\ln(x)\ln^2(1-x)}{1-x}
&=& 2x^2\left[\Sa_{2,2}(x)+\Sa_{1,3}(x) - \ln(x) \Sa_{1,2}(x)
- \frac{5}{4}\zeta_4 \right] 
\\ & &
+3x^2\left[\Sa_{1,2}(x) - \Sa_{1,2}(1-x) - \zeta_3 \right]
\non\\&&
-2x^2[\Li_2(x)-\zeta_2]-\frac{x^2}{2}\Li_2(1-x)
\non\\ & &
+\frac{x^2}{3} \ln(x) \ln^3(1-x) - \frac{1}{2} x^2 [\ln^2(x) + 3 \ln(x)]
 \non\\ & &
+\left[\frac{1}{4} + x - \frac{5}{4} x^2 
+ \left(\frac{1}{2} + x - \frac{3}{2} x^2\right) \ln(x) \right]\ln^2(1-x)
\non\\ & &
+\left[-\frac{3}{2} x(1-x) + \left(\frac{x^2}{2} - x \right) \ln(x)
+ \frac{3}{2} x^2 \ln^2(x) \right] \ln(1-x)
\non\\
x^2 \otimes \frac{\ln^2(x)\ln(1-x)}{1-x}  
&=& 2x^2\left[\left\{\ln(1-x)-\frac{3}{2}\right\}\Sa_{1,2}(1-x) 
- \Sa_{2,2}(1-x) - \Sa_{1,3}(1-x)\right]
\non\\ & &
+ \frac{5}{2} x^2 \Li_2(1-x) 
+ x^2\left(\frac{1}{2}-\frac{1}{3}\ln(1-x)\right)\ln^3(x)
\\ & &
+ \left[-\frac{x}{2}+\frac{5}{4}x^2 
+ \left(\frac{1}{2}+x\right)\ln(1-x)\right]\ln^2(x)
\non\\ & &
+\left[-\frac{3}{2}x + \frac{9}{4}x^2
+ \left(\frac{1}{2}+2x\right)\ln(1-x)\right]\ln(x)
\non\\&&
+ \left(\frac{1}{4}+2x-\frac{9}{4}x^2\right)\ln(1-x)
-\frac{7}{4}x(1-x)
\non\\
x^2 \otimes \frac{\ln^3(x)}{1-x}
&=& -6x^2\Sa_{1,3}(1-x) - \frac{x^2}{4} \ln^4(x) + \left(x+\frac{1}{2}
\right) \ln^3(x) 
\\ & &
+ 3 \left(x+\frac{1}{4}\right) \ln^2(x) 
+ 6\left(x+\frac{1}{8}\right) \ln(x) + \frac{3}{8} + 6x 
- \frac{51}{8} x^2  
\non\\
x^2 \otimes \frac{\ln(x)\Li_2(1-x)}{1-x}
&=& 3x^2\left[\Sa_{1,3}(1-x)+\Sa_{1,2}(1-x) - \frac{1}{6} \Li_2^2(1-x)
\right] 
\\ & &
+\frac{1}{2} \left[-x^2\ln^2(x)+ \left(1+2x\right) \ln(x)
+ \frac{1}{2} + 2x - \frac{5}{2}x^2\right]\Li_2(1-x)
\nonumber\\ & &
-\frac{x^2}{2}\ln^3(x) + \frac{x}{2} \left(1-\frac{5}{4}x\right)
\ln^2(x)+\frac{5}{4} x\ln(x)+ \frac{5}{4} x(1-x)  
\non\\
x^2 \otimes \Li_3(1-x)
&=& \frac{1}{2}\left(1-x^2\right) \Li_3(1-x)
 - \frac{1}{2}\left[x(1-x)-x^2 \ln(x)\right] \Li_2(1-x)
\nonumber\\ & &
+ \frac{x^2}{4} \ln^2(x) + x^2 \Sa_{1,2}(1-x)  
\\
x^2 \otimes x \Li_3(1-x)
&=& x(1-x)\Li_3(1-x) +x^2\ln(x)\Li_2(1-x) +2x^2 \Sa_{1,2}(1-x)  
\\
x^2 \otimes \ln(1-x)\Li_2(1-x)
&=& \frac{3}{2}x^2\Sf(1-x) 
\non\\ & &
+\frac{1}{2}\left[(1-x^2)\ln(1-x)+x^2\ln(x)-x+2x^2\right]\Li_2(1-x)
\\ & &
+ \frac{x}{2}\left[ -\frac{x}{2}\ln^2(x) + \ln(x)
+ 1 - x \right] \ln(1-x) 
\non\\ & &
+ \frac{x^2}{2}[\ln^2(x) + \ln(x)]
\non\\
x^2 \otimes x \ln(1-x)\Li_2(1-x)
&=& x^2\left[2\Sa_{1,2}(1-x) + \ln(x) \Li_2(x) - \Li_3(x) + \zeta_3
\right] \nonumber\\ & &
+x \left[x\ln(x) +(1-x) \ln(1-x) \right] \Li_2(1-x)  
\\
x^2 \otimes \ln(x)\Li_2(1-x) 
&=& x^2 \Sa_{1,2}(1-x) +\frac{1}{2} \left[\ln(x) + \frac{1}{2}\left(
1-x^2\right)\right] \Li_2(1-x) \nonumber\\ & &
- \frac{x^2}{6} \ln^3(x) + \frac{x}{8} (4-x)\ln^2(x) + \frac{5}{4}x
\left[1-x+\ln(x)\right]  
\\
x^2 \otimes x \ln(x)\Li_2(1-x) 
&=& 2 x^2 \Sa_{1,2}(1-x) + x\left[1-x +\ln(x)\right] \Li_2(1-x)
\nonumber\\ & & - x^2 \left[\frac{1}{3} \ln^3(x) + \frac{1}{2} \ln^2(x)
\right]  
\\
x^2 \otimes \Sf(1-x) 
&=& \frac{1}{2}\left(1-x^2\right) \Sa_{1,2}(1-x)
+ \frac{x^2}{12} \ln^3(x) - \frac{x}{4} \ln^2(x) -
\frac{x}{2} \ln(x) \nonumber\\ & &
-\frac{x}{2}(1-x)  
\\
x^2 \otimes x \Sf(1-x)
&=& x(1-x) \Sa_{1,2}(1-x) + \frac{x^2}{6} \ln^3(x)  
\\
x^2 \otimes \ln^3(1-x) 
&=& 
\frac{1}{2} \left(1-x^2\right) \ln^3(1-x) - \frac{3}{2} x
(1-x) \ln^2(1-x) \nonumber\\ & &
+ 3 x^2 \left[\Sa_{1,2}(x)-\zeta_3+\Li_2(x) -\zeta_2\right]   
\\
x^2 \otimes x \ln^3(1-x)
&=& x(1-x) \ln^3(1-x) + 6x^2\left[\Sa_{1,2}(x)-\zeta_3\right]  
\\
x^2 \otimes \ln(x)\ln^2(1-x)
&=& 
x^2\left[\Sa_{1,2}(x)-\Sa_{1,2}(1-x) - \zeta_3\right]
-\frac{x^2}{2}\Li_2(1-x) \nonumber\\ & &
+ \frac{1}{2} \left[\frac{1}{2} + \ln(x) \right] \left(1-x^2\right)
\ln^2(1-x)\nonumber\\ & &
+ \left[-\frac{3}{2}x(1-x) - x \left(1- \frac{x}{2}\right) \ln(x)
+ \frac{x^2}{2}\ln^2(x) \right] \ln(1-x) \nonumber\\ & &
- \frac{x^2}{2} \left[3 \ln(x) + \ln^2(x) \right]  
\\
x^2 \otimes x \ln(x)\ln^2(1-x)
&=& x(1-x)\left[1+\ln(x)\right] \ln^2(1-x) + 2 x^2 \ln(x) \ln(1-x)
\nonumber\\ & & +2x^2\left[\Sa_{1,2}(x) + \Li_3(x) + \Li_2(1-x)
- \ln(x) \Li_2(x) - 2\zeta_3\right]  
\\ 
x^2 \otimes \ln^2(x)\ln(1-x) 
&=& -x^2 \Sa_{1,2}(1-x) + \frac{x^2}{2}\Li_2(1-x) 
\non\\
& & + \left\{\frac{1}{4}\left(1-x^2\right) + \frac{1}{2} \ln(x) \left[
1 + \ln(x) \right]\right\}\ln(1-x) 
\\ & &
+ \frac{x^2}{6} \ln^3(x) - \frac{x}{2}\left(1-\frac{x}{2}\right)
\ln^2(x) - \frac{x}{2}\left(3-\frac{x}{2} \right)\ln(x)
\non\\&&  
- \frac{7}{4}x(1-x)
\non\\
x^2 \otimes x \ln^2(x)\ln(1-x)
&=&  x \left[2(1-x) + \ln^2(x) + 2 \ln(x) \right] \ln(1-x) 
\\& & 
+ 2x^2 \left[\frac{1}{6} \ln^3(x) + \frac{1}{2} \ln^2(x) + \ln(x)
\right]
\nonumber\\ & & + 2 x^2\left[
\Li_2(1-x) - \Sa_{1,2}(1-x) \right]  
\non\\
x^2 \otimes \ln^3(x)  
&=& \frac{1}{2} \ln^3(x) + \frac{3}{4} \ln^2(x) + \frac{3}{4} \ln(x) +
\frac{3}{8} (1-x^2)  
\\
x^2 \otimes x \ln^3(x) 
&=& x \left[\ln^3(x) + 3 \ln^2(x) + 6\ln(x) \right] + 6x(1-x)  
\\
x^2 \Sf(1-x) \otimes \frac{1}{x} 
&=& \frac{1}{3}\left(\frac{1}{x}-x^2\right) \Sa_{1,2}(1-x) + 
\frac{1}{6}\left[1+\frac{x}{2}+\frac{x^2}{3}\right] \ln^2(x)
\\ & &
-\frac{1}{3} \left[1 +\frac{x}{4}+\frac{x^2}{9}\right]\ln(x)
- \frac{251}{648x} + \frac{1}{3} + \frac{x}{24} 
+ \frac{x^2}{81}  
\non\\
x^2 \Li_3(1-x) \otimes \frac{1}{x} 
&=&
\frac{1}{3}\left(\frac{1}{x}-x^2\right)\Li_3(1-x) 
+\left[-\frac{11}{18x}+\frac{1}{3}+\frac{x}{6}+\frac{x^2}{9}\right]\Li_2(1-x)
\nonumber\\ & &
+\left[\frac{11}{18} + \frac{5}{36} x + \frac{x^2}{27}\right] \ln(x)
+\frac{449}{648x} - \frac{11}{18} 
- \frac{5}{72} x - \frac{x^2}{81}  
\\
\frac{1}{x}\ln^2(1-x) \otimes \frac{1}{x}
&=& \frac{2}{x} \left[\zeta_3 - \Sf(x)\right]  
\\
\frac{1}{x}\ln(x) \otimes \frac{1}{x} 
&=& - \frac{1}{2x} \ln^2(x)  
\\
1 \otimes x^2 \Sf(1-x)
&=& \frac{1}{2}\left(1-x^2\right) \Sa_{1,2}(1-x) 
+ \frac{x}{4} \left(1+ \frac{x}{2}\right) \ln^2(x) 
\\ & &
- \frac{x}{2} \left(1 + \frac{x}{4} \right) \ln(x) - \frac{9}{16}
+ \frac{x}{2} + \frac{x^2}{16}  
\non\\
x \otimes x^2 \Sf(1-x)
&=& x(1-x) \left[\Sa_{1,2}(1-x) - 1\right] + x^2\ln(x)\left[
\frac{1}{2} \ln(x) -1 \right]  
\\
1 \otimes x^2 \Li_3(1-x)
&=&  
\frac{1}{2}\left(1-x^2\right) \Li_3(1-x) 
- \frac{1}{4} \left(3-2x-x^2\right) \Li_2(1-x) 
\\ & &
+\frac{x}{4} \left(3+\frac{x}{2}\right) \ln(x) + \frac{13}{16} 
- \frac{3}{4} x - \frac{x^2}{16}  
\non\\
x \otimes x^2 \Li_3(1-x)
&=& 
x(1-x)\left[\Li_3(1-x) - \Li_2(1-x) +1\right] + x^2\ln(x)  
\\
x^2 \otimes x^2 \Li_3(1-x)
&=&
-x^2 \ln(x) \Li_3(1-x) - \frac{x^2}{2} \Li_2^2(1-x)  
\\
1 \otimes \frac{1}{x}\ln^2(1-x) 
&=& \left(\frac{1}{x}-1\right) \ln^2(1-x) - 2\left[\Li_2(x) - \zeta_2\right]
\\
x \otimes \frac{1}{x}\ln^2(1-x)
&=&
\frac{1}{2} \left(\frac{1}{x}-x\right) \ln^2(1-x) -(1-x)\ln(1-x) -x \ln(x)
\nonumber\\ & &
-x\left[\Li_2(x) - \zeta_2\right]  
\\
x^2 \otimes \frac{1}{x}\ln^2(1-x)
&=&
\frac{1}{3} \left(\frac{1}{x}-x^2\right) \ln^2(1-x) - \frac{1}{3}
\left(1+2x-3x^2\right) \ln(1-x) 
\\ & & 
-x^2 \ln(x) 
- \frac{2}{3} x^2 \left[\Li_2(x) - \zeta_2\right] +\frac{1}{3} x(1-x)  
\non\\
1 \otimes \frac{1}{x}\ln(x) 
&=& \frac{1}{x}\left[1-x +\ln(x)\right]  
\\
x \otimes \frac{1}{x}\ln(x)
&=& \frac{1}{2x} \left[\frac{1}{2}(1-x^2) + \ln(x) \right]  
\\
x^2 \otimes \frac{1}{x}\ln(x)
&=& \frac{1}{3x} \left[\frac{1}{3}(1-x^3) + \ln(x)\right]  
\\
x^2\ln(1-x)\Li_2(1-x) \otimes \frac{1}{x}
&=&\left\{-\frac{11}{18x}+\frac{1}{3}+\frac{x}{6}+\frac{x^2}{9}
+\frac{1}{3}\left(\frac{1}{x}-x^2\right)\ln(1-x)\right\}\Li_2(1-x)
\non\\&&
+\frac{11}{18x}[\Li_2(x)-\zeta_2]
+\left(\frac{11}{18x}-\frac{1}{3}
                   -\frac{x}{6}-\frac{x^2}{9}\right)\ln(x)\ln(1-x)
\non\\&&
+\left(-\frac{49}{108x}+\frac{1}{3}
                       +\frac{x}{12}+\frac{x^2}{27}\right)\ln(1-x)
\\&&
+\left(\frac{11}{9}+\frac{5}{18}x+\frac{2}{27}x^2\right)\ln(x)
+\frac{413}{216x}-\frac{181}{108}-\frac{43}{216}x-\frac{x^2}{27}
\non\\
x^2\ln^3(1-x) \otimes \frac{1}{x}
&=& \frac{1}{3}\left(\frac{1}{x}-x^2\right)\ln^3(1-x)
+\left(-\frac{11}{6x}+1+\frac{x}{2}+\frac{x^2}{3}\right)\ln^2(1-x)
\non\\&&
+\left(\frac{85}{18x}-\frac{11}{3}
-\frac{5}{6}x-\frac{2}{9}x^2\right)\ln(1-x)
\\&&
-\frac{575}{108x}+\frac{85}{18}+\frac{19}{36}x+\frac{2}{27}x^2
\non\\
x^2\ln(x)\ln^2(1-x) \otimes \frac{1}{x}
&=& \frac{2}{3x}[\zeta_3-\Sf(x)]
 -\frac{11}{9x}[\Li_2(x)-\zeta_2]
\non\\&&
+\frac{1}{3}\left(\frac{1}{x}-x^2\right)\ln(x)\ln^2(1-x)
-\frac{1}{9}\left(\frac{1}{x}-x^2\right)\ln^2(1-x)
\non\\&&
+\left(-\frac{11}{9x}+\frac{2}{3}
                +\frac{x}{3}+\frac{2}{9}x^2\right)\ln(x)\ln(1-x)
\\&& 
+\left(\frac{71}{54x}-\frac{8}{9}
             -\frac{5}{18}x-\frac{4}{27}x^2\right)\ln(1-x)
\non\\&&
       -\left(\frac{11}{9}+\frac{5}{18}x+\frac{2}{27}x^2\right)\ln(x)
-\frac{80}{27x}+\frac{137}{54}+\frac{19}{54}x +\frac{2}{27}x^2
\non\\
x^2\ln^2(x)\ln(1-x) \otimes \frac{1}{x} 
&=&\frac{2}{3x}\Sf(1-x)+\frac{2}{9x}\Li_2(1-x)
\\&&
+\left(\frac{1}{3}+\frac{x}{6}+\frac{x^2}{9}
                       -\frac{x^2}{3}\ln(1-x)\right)\ln^2(x)
\non\\&&
+\frac{2}{9}x^2\ln(x)\ln(1-x)
-\left(\frac{8}{9}+\frac{5}{18}x+\frac{4}{27}x^2\right)\ln(x)
\non\\&&
+\frac{2}{27}\left(\frac{1}{x}-x^2\right)\ln(1-x)
-\frac{131}{108x}+\frac{26}{27}+\frac{19}{108}x+\frac{2}{27}x^2
\non\\
x^2\ln^3(x) \otimes \frac{1}{x}
&=&-\frac{2}{27}\left(\frac{1}{x}-x^2\right)
-\frac{x^2}{3}\left(\frac{2}{3}\ln(x)-\ln^2(x)+\ln^3(x)\right)
\\
1 \otimes x^2\ln(1-x)\Li_2(1-x)
&=&
\frac{1-x}{2}\left\{(1+x)\ln(1-x)-\frac{3}{2}-\frac{x}{2}\right\}\Li_2(1-x)
\non\\&&
+\frac{3}{4}[\Li_2(x)-\zeta_2]
+\left(\frac{3}{4}-\frac{x}{2}-\frac{x^2}{4}\right)\ln(x)\ln(1-x)
\\&&
+\left(\frac{3}{2}+\frac{x}{4}\right)x\ln(x)
-\frac{1}{8}(5-4x-x^2)\ln(1-x)
\non\\&&
+\frac{37}{16}-\frac{17}{8}x-\frac{3}{16}x^2
\non\\
1 \otimes x^2\ln^3(1-x)
&=&\frac{1}{2}(1-x^2)\ln^3(1-x)
+\left(-\frac{9}{4}+\frac{3}{2}x+\frac{3}{4}x^2\right)\ln^2(1-x)
\\&&
+\left(\frac{21}{4}-\frac{9}{2}x-\frac{3}{4}x^2\right)\ln(1-x)
-\frac{45}{8}+\frac{21}{4}x+\frac{3}{8}x^2
\non\\
1 \otimes x^2\ln(x)\ln^2(1-x)
&=& \zeta_3-\Sf(x)-\frac{3}{2}[\Li_2(x)-\zeta_2]
\\&&
-\frac{1}{4}(1-x^2)[1-2\ln(x)]\ln^2(1-x)
\non\\&&
+\left(-\frac{3}{2}+x+\frac{x^2}{2}\right)\ln(x)\ln(1-x)
+\left(2-\frac{3}{2}x-\frac{x^2}{2}\right)\ln(1-x)
\non\\&&
-\frac{x}{4}(x+6)\ln(x)
-\frac{31}{8}+\frac{7}{2}x+\frac{3}{8}x^2
\non\\
1 \otimes x^2\ln^2(x)\ln(1-x)
&=&\Sf(1-x)+\frac{1}{2}\Li_2(1-x)
\non\\&&
+\frac{x}{2}[1+\frac{x}{2}-x\ln(1-x)]\ln^2(x)
\\&&
-\frac{x}{2}[3+x-x\ln(1-x)]\ln(x)
\non\\&&
+\frac{1}{4}(1-x^2)\ln(1-x)
-\frac{17}{8}+\frac{7}{4}x+\frac{3}{8}x^2
\non\\
1 \otimes x^2\ln^3(x)
&=&-\frac{x^2}{2}
\left\{\ln^3(x)-\frac{3}{2}\ln^2(x)+\frac{3}{2}\ln(x)\right\}
-\frac{3}{8}(1-x^2)
\\
x \otimes x^2\ln(1-x)\Li_2(1-x)
&=& -x\Li_2(1-x)+x(1-x)[\ln(1-x)-1]\Li_2(1-x)
\\&&
+x(1-x)[3-\ln(1-x)]+x^2\ln(x)[2-\ln(1-x)]
\non\\
x \otimes x^2\ln^3(1 - x)
&=& -x(1-x)[6-6\ln(1-x)+3\ln^2(1-x)-\ln^3(1-x)]
\\
x \otimes x^2\ln(x)\ln^2(1-x)
&=& 2x[\zeta_3-\Sf(x)]+2x[\zeta_2-\Li_2(x)]-2x\ln(x)
\\&&
-x(1-x)[6-4\ln(1-x)+\ln^2(1-x)]
\non\\&&
+x(1-x)\ln(x)[2-2\ln(1-x)+\ln^2(1-x)]
\non\\
x \otimes x^2\ln^2(x)\ln(1-x)
&=& 2x[\Sf(1-x)+\Li_2(1-x)]
\\&&
+x^2\ln(x)\{-2+[2-\ln(x)][\ln(1-x)-1]\}
\non\\&&
+2x(1-x)[\ln(1-x)-3]
\non\\
x \otimes x^2\ln^3(x)
&=& -x^2\ln^3(x)+3x^2\ln^2(x)-6x(x\ln(x)-x+1)
\\
x^2 \otimes x^2\ln^2(1-x)
&=& 2 x^2 [ \zeta(3) - \Sf(x) ]
\\
x^2 \otimes x^2\ln(x)\ln(1-x)
&=& x^2[\zeta_3+\ln(x)\Li_2(x)-\Li_3(x)]
\\
x^2 \otimes x^2\ln^2(x)
&=& -\frac{x^2}{3}\ln^3(x)
\\
x^2 \otimes x^2\Li_2(1-x)
&=& -x^2[2\Sf(1-x)+\ln(x)\Li_2(1-x)]
\\
x^2 \otimes \frac{\Sf(1-x)}{1-x}
&=& x^2[\SS_{2,2}(1-x)-3\SS_{1,3}(1-x)]
\\&&
+\left[-x^2\ln(x)+\frac{1}{2}+x-\frac{3}{2}x^2\right]\Sf(1-x)
\non\\&&
+\frac{x^2}{4}\ln^3(x)-\frac{x}{4}\ln^2(x)-\frac{x}{2}\ln(x)
-\frac{1}{2}x(1-x)
\non
\end{eqnarray}

\vspace{3mm}
\noindent
{\bf Acknowledgments:} This work was supported in part by DFG
Sonderforschungsbereich Transregio 9, Computergest\"utzte Theoretische
Physik.

\newpage


\begin{thebibliography}{99}
%
\bibitem{EXAMP1}
see e.g.~: A. Akhundov,
D. Bardin, and T. Riemann, Nucl. Phys. {\bf B276} (1986) 1;\\
S. Jadach and  B. Ward, Comput. Phys. Commun. {\bf 56} (1990) 351, {\bf 66} (1991) 
276; {\bf 124} (2000) 233;\\
M. Consoli and M. Greco, Nucl. Phys. {\bf B186} (1981) 519.
%
\bibitem{Beenakker:1989km}
  W.~Beenakker, F.~A.~Berends and W.~L.~van Neerven,
Print-89-0445 (Leiden)
Proc. Workshop on Electroweak Radiative Corrections, Ringberg Castle, Germany, 
Apr 3-7, 1989, p.~3.
%
\bibitem{Blumlein:1989gk}
  J.~Bl\"umlein,
  Z.\ Phys.\  {\bf C47} (1990) 89.
%
\bibitem{Bardin:1996ch}
  D.~Y.~Bardin, J.~Bl\"umlein, P.~Christova and L.~Kalinovskaya,
  Nucl.\ Phys.\ {\bf B 506} (1997) 295
  [arXiv:hep-ph/9612435].
%
\bibitem{ILC}
  E.~Accomando {\it et al.}  [ECFA/DESY LC Physics Working Group],
  Phys.\ Rept.\  {\bf 299} (1998) 1
  [arXiv:hep-ph/9705442];\\
  J.~A.~Aguilar-Saavedra {\it et al.}  [ECFA/DESY LC Physics Working Group],
  arXiv:hep-ph/0106315.
%
\bibitem{Berends:1987ab}
F.~A.~Berends, W.~L.~van Neerven and G.~J.~Burgers,
Nucl.\ Phys.\  {\bf B297} (1988) 429
[Erratum-ibid.\  {\bf B304} (1988) 921].
%
\bibitem{Altarelli:1986kq}
  G.~Altarelli and G.~Martinelli,
  in J. Ellis, R.D. Peccei ( Eds.): Physics At LEP, Vol. {\bf 1}, p~47. 
%
\bibitem{Blumlein:1994ii}
  J.~Bl\"umlein,
  Z.\ Phys.\  {\bf C65} (1995) 293
  [arXiv:hep-ph/9403342].
%
\bibitem{Arbuzov:1995id}
  A.~Arbuzov, D.~Y.~Bardin, J.~Bl\"umlein, L.~Kalinovskaya and T.~Riemann,
  Comput.\ Phys.\ Commun.\  {\bf 94} (1996) 128
  [arXiv:hep-ph/9511434].
%
\bibitem{Jezabek:1992bx}
M.~Jezabek,
Z.\ Phys.\  {\bf C56} (1992) 285.
%
\bibitem{Montagna:1996jv}
  G.~Montagna, O.~Nicrosini and F.~Piccinini,
  Phys.\ Lett.\  {\bf B406} (1997) 243
  [arXiv:hep-ph/9611463].
%
\bibitem{Skrzypek:1992vk}
M.~Skrzypek,
Acta Phys.\ Polon.\  {\bf B23} (1992) 135.
%
\bibitem{Przybycien:1993qe}
M.~Przybycien,
Acta Phys.\ Polon.\  {\bf B24} (1993) 1105 [arXiv:hep-th/9511029].
%
\bibitem{Arbuzov:1999cq}
A.~B.~Arbuzov,
Phys.\ Lett.\  {\bf B470} (1999) 252 [arXiv:hep-ph/9908361].
%
\bibitem{Blumlein:2004bs}
  J.~Bl\"umlein and H.~Kawamura,
  Nucl.\ Phys.\  {\bf B708} (2005) 467
  [arXiv:hep-ph/0409289].
%
\bibitem{BLKA}
  J.~Bl\"umlein and H.~Kawamura,
  Acta Phys.\ Polon.\  {\bf B33} (2002) 3719
  [arXiv:hep-ph/0207259];
  [arXiv:hep-ph/0309135];
  Nucl.\ Phys.\ Proc.\ Suppl.\  {\bf 116} (2003) 110
  [arXiv:hep-ph/0211219].
%
\bibitem{Blumlein:1998yz}
J.~Bl\"umlein, S.~Riemersma and A.~Vogt,
Eur.\ Phys.\ J.\  {\bf C1} (1998) 255 [arXiv:hep-ph/9611214];
  Nucl.\ Phys.\ Proc.\ Suppl.\  {\bf 51C} (1996) 30
  [arXiv:hep-ph/9608470].
%
\bibitem{SX1}
R.~Kirschner and L.~N.~Lipatov,
Nucl.\ Phys.\ {\bf B213} (1983) 122 
;\\
J.~Bl\"umlein and A.~Vogt,
Phys.\ Lett.\  {\bf B370} (1996) 149
[arXiv:hep-ph/9510410];
Phys.\ Lett.\ {\bf B386} (1996) 350
[arXiv:hep-ph/9606254];
Acta Phys.\ Polon.\  {\bf B27} (1996) 1309
[arXiv:hep-ph/9603450];\\
J.~Bartels, B.~I.~Ermolaev and M.~G.~Ryskin,
Z.\ Phys.\  {\bf C72} (1996) 627 [arXiv:hep-ph/9603204];\\
  Y.~Kiyo, J.~Kodaira and H.~Tochimura,
  Z.\ Phys.\  {\bf C74} (1997) 631
  [arXiv:hep-ph/9701365].
%
\bibitem{Fadin:1975cb}
  V.~S.~Fadin, E.~A.~Kuraev and L.~N.~Lipatov,
  Phys.\ Lett.\  {\bf B60} (1975) 50.
%
\bibitem{TAR}
  O.~V.~Tarasov, A.~A.~Vladimirov and A.~Y.~Zharkov,
  Phys.\ Lett.\  {\bf B93} (1980) 429;\\
  S.~A.~Larin and J.~A.~M.~Vermaseren,
  Phys.\ Lett.\  {\bf B303} (1993) 334
  [arXiv:hep-ph/9302208].
%
\bibitem{Furmanski:1981cw}
  W.~Furmanski and R.~Petronzio,
  Z.\ Phys.\  {\bf C11} (1982) 293.
%
\bibitem{Blumlein:1997em}
  J.~Bl\"umlein and A.~Vogt,
  Phys.\ Rev.\  {\bf D58} (1998) 014020
  [arXiv:hep-ph/9712546].
%
\bibitem{Blumlein:2002fy}
  J.~Bl\"umlein and H.~Kawamura,
  Phys.\ Lett.\  {\bf B553} (2003) 242
  [arXiv:hep-ph/0211191].
%
\bibitem{Arbuzov:2002cn}
  A.~Arbuzov and K.~Melnikov,
  Phys.\ Rev.\  {\bf D66} (2002) 093003
  [arXiv:hep-ph/0205172].
%
\bibitem{Blumlein:2000wh}
  J.~Bl\"umlein, V.~Ravindran and W.~L.~van Neerven,
  Nucl.\ Phys.\  {\bf B586} (2000) 349
  [arXiv:hep-ph/0004172].
%
\bibitem{Gross:1974cs}
  D.~J.~Gross and F.~Wilczek,
  Phys.\ Rev.\  {\bf D9} (1974) 980.
%
\bibitem{Altarelli:1977zs}
  G.~Altarelli and G.~Parisi,
  Nucl.\ Phys.\  {\bf B126} (1977) 298.
%
\bibitem{DeRujula:1979jj}
  A.~De Rujula, R.~Petronzio and A.~Savoy-Navarro,
  Nucl.\ Phys.\  {\bf B154} (1979) 394.
%
\bibitem{Kuraev:1985hb}
  E.~A.~Kuraev and V.~S.~Fadin,
  Sov.\ J.\ Nucl.\ Phys.\  {\bf 41} (1985) 466
  [Yad.\ Fiz.\  {\bf 41} (1985) 733].
%
\bibitem{Mo:1968cg}
  L.~W.~Mo and Y.~S.~Tsai,
  Rev.\ Mod.\ Phys.\  {\bf 41} (1969) 205.
%
\bibitem{Blumlein:1993ef}
  J.~Bl\"umlein, G.~Levman and H.~Spiesberger,
  J.\ Phys.\  {\bf G19} (1993) 1695.
%
\bibitem{NLO}
  W.~Furmanski and R.~Petronzio,
  Phys.\ Lett.\ {\bf B97} (1980) 437;\\
  E.~G.~Floratos, C.~Kounnas and R.~Lacaze,
  Nucl.\ Phys.\  {\bf B192} (1981) 417;\\
  W.~K.~Tung,
  Nucl.\ Phys.\  {\bf B315} (1989) 378.
%
\bibitem{QCDVA}
  J.~Kripfganz and H.~Perlt,
  Z.\ Phys.\  {\bf C41} (1988) 319;\\
  H.~Spiesberger,
  Phys.\ Rev.\  {\bf D52} (1995) 4936
  [arXiv:hep-ph/9412286];\\
  A.~D.~Martin, R.~G.~Roberts, W.~J.~Stirling and R.~S.~Thorne,
  Eur.\ Phys.\ J.\  {\bf C39} (2005) 155
  [arXiv:hep-ph/0411040].
%
\bibitem{LQ}
  J.~Ohnemus, S.~Rudaz, T.~F.~Walsh and P.~M.~Zerwas,
  Phys.\ Lett.\  {\bf B334} (1994) 203
  [arXiv:hep-ph/9406235];\\
  J.~Bl\"umlein and E.~Boos,
  Nucl.\ Phys.\ Proc.\ Suppl.\  {\bf 37B} (1994) 181.
%
\bibitem{Blumlein:2003gb}
  J.~Bl\"umlein,
  Comput.\ Phys.\ Commun.\  {\bf 159} (2004) 19
  [arXiv:hep-ph/0311046].
%
\bibitem{Blumlein:1999if}
J.~Bl\"umlein and S.~Kurth,
Phys.\ Rev.\  {\bf D60} (1999) 014018 [arXiv:hep-ph/9810241].
\end{thebibliography}
\end{document}